\documentclass[12pt]{spieman}  
\usepackage{amsmath,amsfonts,amssymb}
\usepackage{graphicx}
\usepackage{setspace}
\usepackage{tocloft}
\usepackage[colorlinks=true, allcolors=blue]{hyperref}
\usepackage{svg}
\usepackage{subfig}
\usepackage{lineno}
\linenumbers

\usepackage{xcolor}
\newcommand\todo[1]{\textcolor{black}{#1}}
\newcommand\todos[1]{\textcolor{black}{#1}}

\title{The WALOP-North Instrument I: Optical Design, Filter Design, Calibration }

\author[a,b*]{John A. Kypriotakis}
\author[c, d*]{Siddharth Maharana}
\author[c,j]{Ramya M. Anche}
\author[c]{Chaitanya V. Rajarshi}
\author[c,a,e]{A. N. Ramaprakash}
\author[c]{Bhushan Joshi}
\author[g]{Artem Basyrov}
\author[a,b]{Dmitry Blinov}
\author[h]{Tuhin Ghosh}
\author[g]{Eirik Gjerløw}
\author[a,b]{Sebastian Kiehlmann}
\author[a,b]{Nikolaos Mandarakas}
\author[f]{Georgia V. Panopoulou}
\author[a,b]{Katerina Papadaki}
\author[a,b]{Vasiliki Pavlidou}
\author[e]{Timothy J. Pearson}
\author[k]{Vincent Pelgrims}
\author[d,i]{Stephen B. Potter}
\author[e,l]{Anthony C. S. Readhead}
\author[l]{Raphael Skalidis}
\author[a,b]{Konstantinos Tassis}
\affil[a]{Institute of Astrophysics, Foundation for Research and Technology-Hellas, Voutes, 70013 Heraklion, Greece}
\affil[b]{Department of Physics, University of Crete, Voutes, 70013 Heraklion, Greece}
\affil[c]{Inter-University Centre for Astronomy and Astrophysics, Post bag 4, Ganeshkhind, Pune, 411007, India}
\affil[d]{South African Astronomical Observatory, PO Box 9, Observatory, 7935, Cape Town, South Africa}
\affil[e]{Cahill Center for Astronomy and Astrophysics, California Institute of Technology, Pasadena, CA, 91125, USA}
\affil[f]{Department of Space, Earth and Environment, Chalmers University of Technology, 412 93, Gothenburg, Sweden}
\affil[g]{Institute of Theoretical Astrophysics, University of Oslo, P.O. Box 1029 Blindern, NO-0315 Oslo, Norway}
\affil[h]{School of Physical Sciences, National Institute of Science Education and Research, HBNI, Jatni 752050, Odisha, India}
\affil[i]{Department of Physics, University of Johannesburg, PO Box 524, Auckland Park 2006, South Africa}
\affil[j]{Steward Observatory, University of Arizona, Tucson, Arizona, 85721, USA}
\affil[k]{Universit\'e Libre de Bruxelles, Science Faculty CP230, B-1050 Brussels, Belgium}
\affil[l]{Owens Valley Radio Observatory, California Institute of Technology, MC 249-17, Pasadena, CA 91125, USA}

\cftpagenumbersoff{figure}
\cftpagenumbersoff{table} 
\begin{document} 
\maketitle

\begin{abstract}
The Wide Area Linear Optical Polarimeter North (WALOP-North) is an optical polarimeter designed for the needs of the PASIPHAE survey. It will be installed on the 1.3m telescope at the Skinakas Observatory in Crete, Greece. After commissioning, it will measure the polarization of millions of stars at high Galactic latitude, aiming to measure hundreds of stars per $deg^2$. The astronomical filter used in the instrument is a modified, polarimetrically-neutral broadband SDSS-r. This instrument will be pioneering one due to its large field-of-view (FoV) of $30\times 30$ $arcmin^2$ and high accuracy polarimetry measurements. The accuracy and sensitivity of the instrument in polarization fraction will be at the 0.1\% and 0.05\% level, respectively. Four separate 4k$\times$4k CCDs will be used as the instrument detectors, each imaging one of the $0\deg{}, 45\deg{}, 90\deg{}$ and $135\deg{}$ polarized FoV separately, therefore making the instrument a four-channel, one-shot polarimeter. Here, we present the overall optical design of the instrument, emphasizing on the aspects of the instrument that are different from WALOP-South. \todo{We also present a novel design of filters appropriate for polarimetry along with details on the management of the instrument size and its polarimetric calibration.}
\end{abstract}


{\noindent \footnotesize\textbf{*}J.A.K.: ikypriot@physics.uoc.gr\\  \textbf{*}S.M.: siddharth@saao.ac.za }

\begin{spacing}{2}   

\section{Introduction}\label{sect:intro}
Our collaboration, PASIPHAE (Polar-Areas Stellar Imaging in Polarimetry High Accuracy Experiment)\cite{pasiwhite}, aims to map the magnetic field permeating the Galaxy's dust clouds in three dimensions by analyzing the polarization of stars. Starlight is usually not intrinsically polarized, but when there is interstellar dust aligned with the magnetic field in the path between the observer and the star, the light becomes polarized. The presence of this Galactic foreground hinders accurate measurements of the polarization of the Cosmic Microwave Background. By inferring the direction of the Galactic magnetic field in different interstellar clouds along the line of sight, using stars as probes, the PASIPHAE survey seeks to remedy that \cite{pasiwhite,vincent_decorrelation_paper, vincent_tomography}. Wide field polarimeters with high measurement accuracy (up to 0.1\% in polarization fraction) are needed to achieve this goal\cite{pasiwhite}. Two dedicated instruments, Wide Area Liner Optical Polarimeters (WALOP-North and WALOP-South), are constructed in IUCAA, India. The survey will cover high latitude regions in both Galactic hemispheres after these two instruments are commissioned at the Skinakas Observatory in Greece (WALOP-North) and the SAAO Observatory in South Africa (WALOP-South), respectively. Of the two WALOP instruments, WALOP-South\cite{walop_s_spie_2020, SouthOptical, WALOP_Calibration_paper} was designed first and used as a reference to make the WALOP-North instrument. In this article, we discuss the optical design and calibration of WALOP-North, focusing on aspects of the design that are different from that of  WALOP-South. Section \ref{sec:techreq} presents the technical requirements from the instrument design. Section \ref{sec:instdesign} gives the details of the instrument design. Section \ref{sec:perf} presents the instrument performance. Section \ref{sec:filt} documents the procedure of choosing the filters and designing ones appropriate for polarimetry. Finally, section \ref{sec:calib} presents the polarimetric calibration strategy for the instrument.

\section{Technical Requirements}\label{sec:techreq}
Table \ref{tab:techreq} lists the technical specifications for the instrument that stem from PASIPHAE survey's scientific goals \cite{pasiwhite}. These choices were made after considering the materials and technology readily available, as well as the most recent advancements in polarimeter designs.

\subsection{Sensitivity and accuracy}
\todo{The science requirements demand that the device be able to obtain a \todos{$5\sigma$} measurement (before bias \todos{correction}\cite{Wardle,simmons}) for $p\ge0.5\%$ and magnitude $\le16.5$ in the sdss-r' band.} In other words, the instrument's sensitivity (\textit{s}) and accuracy (\textit{a}) must be better than 0.05\% and 0.1\%, respectively. We define \textit{s} as the maximum systematic uncertainty and internal noise introduced by the instrument, and \textit{a} as the largest standard deviation \todos{(STD)} of the measured polarization from the actual polarization respectively (for a star of magnitude 16.5 and polarization fraction 0.5 with an exposure of 20 minutes).

The calculations mentioned above refer to the stars' ideal photometry. For it to be possible, the instrument's imaging capabilities must be superior to (or at least on par with) the site's seeing\cite{skisky}. This means that the instrument must accurately sample the stars' point spread function (PSF) while minimizing any PSF distortion or enlargement caused by air turbulence.

\begin{table}[!ht]
\begin{center}       
\begin{tabular}{|c|c|} 
\hline
\rule[-1ex]{0pt}{3.5ex}  Minimum Polarimetric Sensitivity & 0.05\%  \\
\hline
\rule[-1ex]{0pt}{3.5ex}  Minimum Polarimetric Accuracy & 0.1\%  \\
\hline
\rule[-1ex]{0pt}{3.5ex}  FoV & $30\times{}30$ $arcmin^2$  \\
\hline
\rule[-1ex]{0pt}{3.5ex}  Shots per Measurement & 1   \\
\hline
\rule[-1ex]{0pt}{3.5ex}  Channels (\# CCDs) & 4   \\
\hline
\rule[-1ex]{0pt}{3.5ex}  Imaging Performance & seeing limited PSF size ($1.1 arcsec$ FWHM)  \\
\hline
\rule[-1ex]{0pt}{3.5ex}  CCD Size (px) & $4096\times4096$  \\
\hline
\rule[-1ex]{0pt}{3.5ex}  Pixel Size & $15\mu{}m\times{}15\mu{}m$  \\
\hline
\rule[-1ex]{0pt}{3.5ex}  Main Optical Filter & SDSS-r  \\
\hline
\rule[-1ex]{0pt}{3.5ex}  Stray\&Ghost Light Level & Least possible, less than the Sky brightness \\
\hline
\rule[-1ex]{0pt}{3.5ex}  Size & Compatible with the Skinakas 1.3 m telescope  \\
\hline
\end{tabular}
\caption{\label{tab:techreq}Technical specifications for the WALOP-North Instrument.} 
\end{center}
\end{table} 

\subsection{Field of view and detectors}
For the initial survey to be completed in 2 years, the instrument's field of view (FoV) must be around 30'$\times$30', per the technical requirements and accuracy determined above. \todo{In addition, the CCDs must have 4096 pixels in each dimension, therefore sampling the sky with a plate-scale of $0.43945\frac{arcsec}{px}$, in order to accurately sample the PSF (at least 2 pixels per full width at half maximum (FWHM) - Nyquist limit\cite{Nyquist, Shannon}) at Skinakas (mean seeing FWHM of 1.1'').} In this setting, the instrument's plate scale decreases to 0.44'' per pixel, enabling the PSF to occupy 2.3 pixels per fwhm. Experience has proven that one-shot polarimetry is preferable for achieving the accuracy and sensitivity requirements mentioned earlier, in the shortest amount of time possible. Such polarimeters, like RoboPol \cite{robopol_instrument}, measure the Stokes parameters precisely and in a timely fashion. In this kind of polarimetry, images of the target are concurrently captured at polarization angles of 0, 45, 90, and 135 degrees. \todo{The improved accuracy of 4-channel polarimeters comes from the fact that the signal is simultaneously acquired in all needed channels, therefore reducing what would be channel-to-channel variability (e.g. due to atmospheric effects on the channels throughput).} \todo{These polarimeters do suffer from differential effects across the multiple beams and either had a narrow field of view or lower accuracy over a wider field\cite{hippi, dipol} in an attempt to mitigate those differential effects}. In our case, sacrificing either the FoV size or accuracy was unacceptable due to the survey's time constraints and the instrument's accuracy restrictions. As a result, we needed to use four different, \todo{yet extremely precise} CCDs, each for imaging a different polarization angle from the ones mentioned above (four channels, one shot polarimetry). 

\subsection{Size of the instrument}
\begin{figure}[!ht]
\begin{center}
\begin{tabular}{c}
\includegraphics[height=8cm]{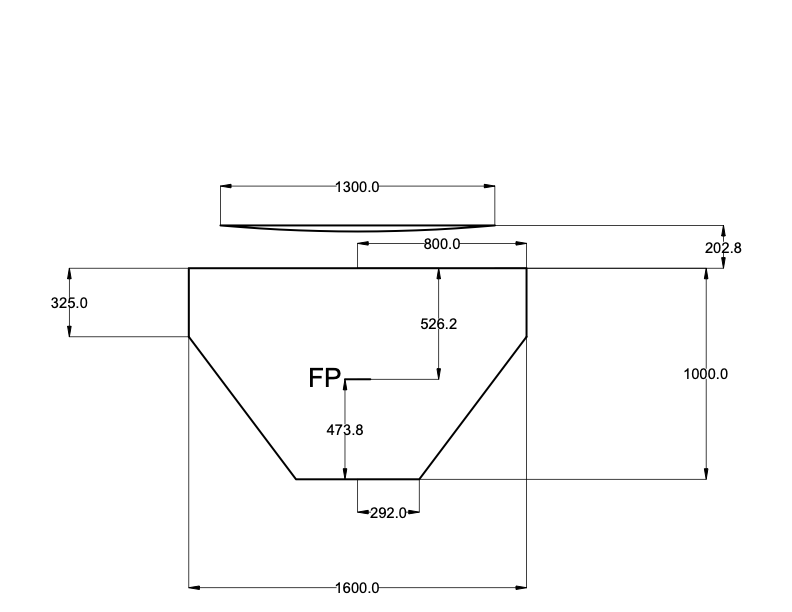}
\end{tabular}
\end{center}
\caption 
{ \label{fig:clearance}
The clearance inside the fork of the Skinakas observatory's 1.3m telescope. All measurements are in mm. FP is the used telescope's focal plane, after the shift described in Section \ref{subsec:teldetails}.} 
\end{figure}

The instrument will be mounted on the 1.3-meter telescope at the Skinakas observatory. The weight of the instrument must not be greater than roughly 200 kg in order not to exceed the torque limit of the telescope's motors. The instrument's size is another constraint because it must fit inside the telescope's mounting fork. In Figure \ref{fig:clearance}, the fork clearance is displayed. The focal plane (FP) is separated from the top of the fork body by a distance of 473.8 mm, and from the back of the primary mirror cell by a distance of 526.2 mm. This permits an instrument to have a maximum length of 1000 mm, assuming that the optics will be positioned on the optical axis before the focal plane.

\subsection{Filter}
Ghost rays and stray light will affect such an instrument, by introducing extra background intensity and skewing the morphology of the image features. It is up to the design to reduce them to the brightness of the sky (or less), effectively eliminating their impact on the measurement. The instrument must function with the SDSS-r filter in place (Section \ref{sec:filt}). This necessitates that it be tuned for wavelengths in the $5000$–$7000$\AA{} range, while no polarization must be added by the filters utilized.

\section{Instrument Design}\label{sec:instdesign}
\begin{figure}[!ht]
\begin{center}
\begin{tabular}{c}
\includegraphics[width=0.9\textwidth]{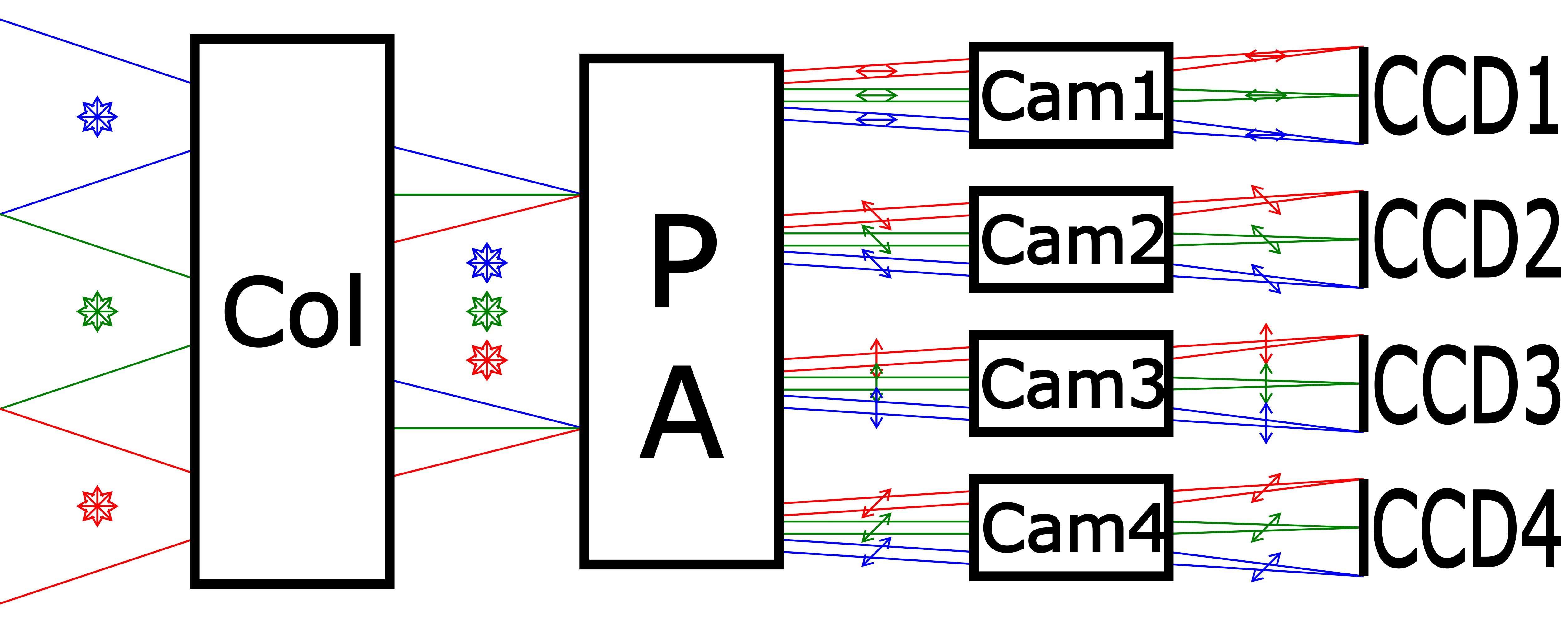}
\end{tabular}
\end{center}
\caption 
{ \label{fig:concept}
\todo{A conceptual diagram of the WALOP instruments. Light enters the collimator ("Col" - from the left, coming from the telescope). It is then fed to the polarization array ("PA" - containing 2 Wollaston prisms side-by-side each preceded by a HWP), whereby is split into 4 beams. The now polarized light is focused by 4 camera lens arrays ("Cam1-4") onto 4 separate CCDs. The diagram represents the WALOP-North instrument whose collimator receives unfocused light (as discussed in Section \ref{sec:instdesign}), while WALOP-South receives the light already focused by the telescope. The arrows denote the polarization of the rays.}} 
\end{figure}

\begin{figure}[!ht]
\begin{center}
\includegraphics[height=15cm]{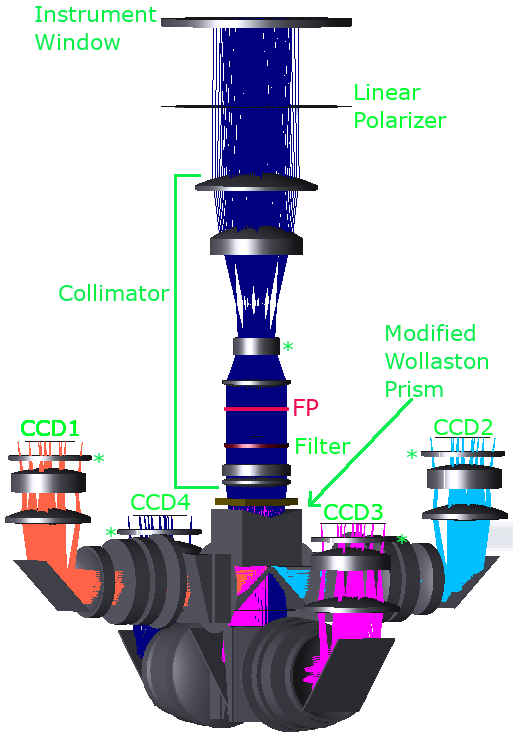}
\end{center}
\caption 
{ \label{fig:shaded}
\todo{The shaded model of the instrument. Light coming from above passes through the collimator, enters the side-by-side Wollaston prisms (being split in 4 beams by polarization) and is re-directed horizontally to 4 separate camera arms. In each of those arms, a second fold redirects the light upwards towards a CCD. The linear polarizer is removable and rotatable, to be used for calibration (Section \ref{sec:calib}). Asterisks denote aspheric lenses. FP is the focal plane of the telescope (without the instrument optics) in the modified configuration of Section \ref{subsec:teldetails}.}}
\end{figure}

\todo{This section will discuss the instrument design and the challenges encountered to attain the desired performance. The design concept of both WALOP-North and its couterpart WALOP-South is presented in Figure \ref{fig:concept}, while the annotated shaded model of the instrument is shown in Figure \ref{fig:shaded}.}

WALOP-North's optical design is similar to the WALOP-South instrument\cite{SouthOptical, walop_s_spie_2020, WALOP_Calibration_paper} ,  as identical polarizer optics are used; the differences are in the design of the collimator and cameras. This paper will concentrate on the key differences between the instruments listed below:
\begin{itemize}
\item Target Telescope
\item Optics Placement
\item Folding of the Beams
\item Usage of Aspherical Optics
\item Guider Design
\end{itemize}

Other minor differences between the two instruments include the placement of the filter between collimator lenses (in the case of WALOP-North) and the size of the FoV that, in the case of WALOP-South, is increased to 35'x35'.

\subsection{Telescope and Site Details}\label{subsec:teldetails}
WALOP-North will operate at the 1.3m telescope at the Skinakas observatory. The instrument, which is directly affixed to the telescope's chassis structure, will take the place of the current GAM (Guiding and Acquisition Module).

In order to meet the size requirements for the instrument, we decided to shift the focal plane of the instrument upwards (closer to the primary mirror) by $119.8mm$. \todo{In order to do this we have to move the secondary mirror from the  nominal position by about $15mm$.} This shifted the f-number of the telescope from $f/7.6$ to $f/8$. \todo{The shifted focus is not the nominal for the telescope, nor the optimal in terms of spot sizes produced at the new focal plane. The collimator lenses of WALOP-North correct the spot sizes (acting as pre-optics in addition to their collimating duty). This is a major difference from the WALOP-South design.} 

\todo{Table \ref{tab:teldetails} lists some details of the telescope and observatory that will host the instrument. These were used in the design of WALOP-North to guide the process and check the instrument compatibility and performance.}

\begin{table}[!ht]
\begin{center}       
\begin{tabular}{|c|c|} 
\hline
\rule[-1ex]{0pt}{3.5ex}  Telescope Type & Modified Ritchey-Chr\'{e}tien  \\
\hline
\rule[-1ex]{0pt}{3.5ex}  Primary Mirror Diameter & 1290mm  \\
\hline
\rule[-1ex]{0pt}{3.5ex}  Secondary Mirror Diameter & 429mm  \\
\hline
\rule[-1ex]{0pt}{3.5ex}  Telescope's Nominal f/\# & 7.6   \\
\hline
\rule[-1ex]{0pt}{3.5ex}  Telescope's Used f/\# & 8   \\
\hline
\rule[-1ex]{0pt}{3.5ex}  \todo{Plate Scale at the focal plane (FP)} & 0.01991$\frac{arcsec}{\mu{}m}$    \\
\hline
\rule[-1ex]{0pt}{3.5ex}  \todos{Plate Scale of instrument + telescope} & 0.02930$\frac{arcsec}{\mu{}m}$    \\
\hline
\rule[-1ex]{0pt}{3.5ex}  \todos{Pixel Size of the WALOP CCDs} & $15\mu{}m\times{}15\mu{}m$  \\
\hline
\rule[-1ex]{0pt}{3.5ex}  Optical Axis Between FP and Chassis (upwards) & 526.2mm \\
\hline
\rule[-1ex]{0pt}{3.5ex}  Optical Axis Between FP and Chassis (downwards) & 473.8mm  \\
\hline
\rule[-1ex]{0pt}{3.5ex}  \todo{Site Altitude (above Mean Sea Level)} & 1750m  \\
\hline
\rule[-1ex]{0pt}{3.5ex}  Median Seeing of the Site & 1.1" \\
\hline
\rule[-1ex]{0pt}{3.5ex}  Operational Temperature Range at the Site & (-5 - 30)$^{\circ{}}$C \\
\hline
\end{tabular}
\caption{\label{tab:teldetails}Telescope and Site Details for the Skinakas 1.3m telescope.} 
\end{center}
\end{table}

\subsection{Optics Placement}\label{subsec:optplace}
In the case of WALOP-North, we had to place some of the collimator optics before the focal plane of the telescope (pre-optics) due to size limitations. This enabled us to save space along the optical (z-) axis of the instrument, therefore allowing us to match the constraints.

\subsection{Folding of the Beams}\label{subsec:fold}
Another strategy used in WALOP-North exclusively was folding the beams an additional time within the camera optics, as seen in Figure \ref{fig:shaded}. \todo{The camera fold happens towards the negative-z direction to \todo{avoid reduction of transmittance due to} Brewster-angle effects~\cite{brewster}.} This was a necessity since the size constraints along the $x$ and $y$ axes were not permitting of a larger instrument.

\subsection{Usage of Aspherics}\label{subsec:asph}
One of the instruments' design goals was to minimize the usage of aspheric lenses in the design due to the complications arising from their utilization. One such complication is that aspheric lenses are much harder to manufacture, and therefore more expensive and perhaps unfeasible. Another hurdle in the integration of aspheric lenses is their proper alignment with the optical beam. Due to the space constraints in WALOP-North, this type of lens was not avoided completely, and one such lens exists in the collimator, in addition to one more aspheric lens in each camera arm. Details of said lenses are given in Table~\ref{tab:asph}.

\begin{table}[ht]
\begin{center}       
\begin{tabular}{|c|c|c|c|c|c|} 
\hline
\rule[-1ex]{0pt}{3.5ex}  \textbf{Lens} & \textbf{ROC} & \textbf{Glass} & \textbf{t} & \textbf{D} & \textbf{k}  \\
\hline
\hline
\rule[-1ex]{0pt}{3.5ex}  Collimator, Lens 3 & -50/98 & S-FPL53 & 10mm & 53mm & 0.253  \\
\hline
\rule[-1ex]{0pt}{3.5ex}  Ordinary Cameras, Lens 6 & -62/-171 & H-QF50 & 15.8mm & 90.4mm & -1.262  \\
\hline
\rule[-1ex]{0pt}{3.5ex}  Extraordinary Cameras, Lens 6 & -62/-171 & H-QF50 & 15.8mm & 90.4mm & -1.330   \\
\hline
\end{tabular}
\caption{\label{tab:asph}Aspherical Lenses used in WALOP-North. ROC is the radius of curvature, D is the lens diameter, t is the thickness, and k is the conic parameter} 
\end{center}
\end{table}

\subsection{Guider Design}\label{subsec:guider}
\begin{figure}[!ht]
\begin{center}
\begin{tabular}{c}
\includegraphics[height=9cm]{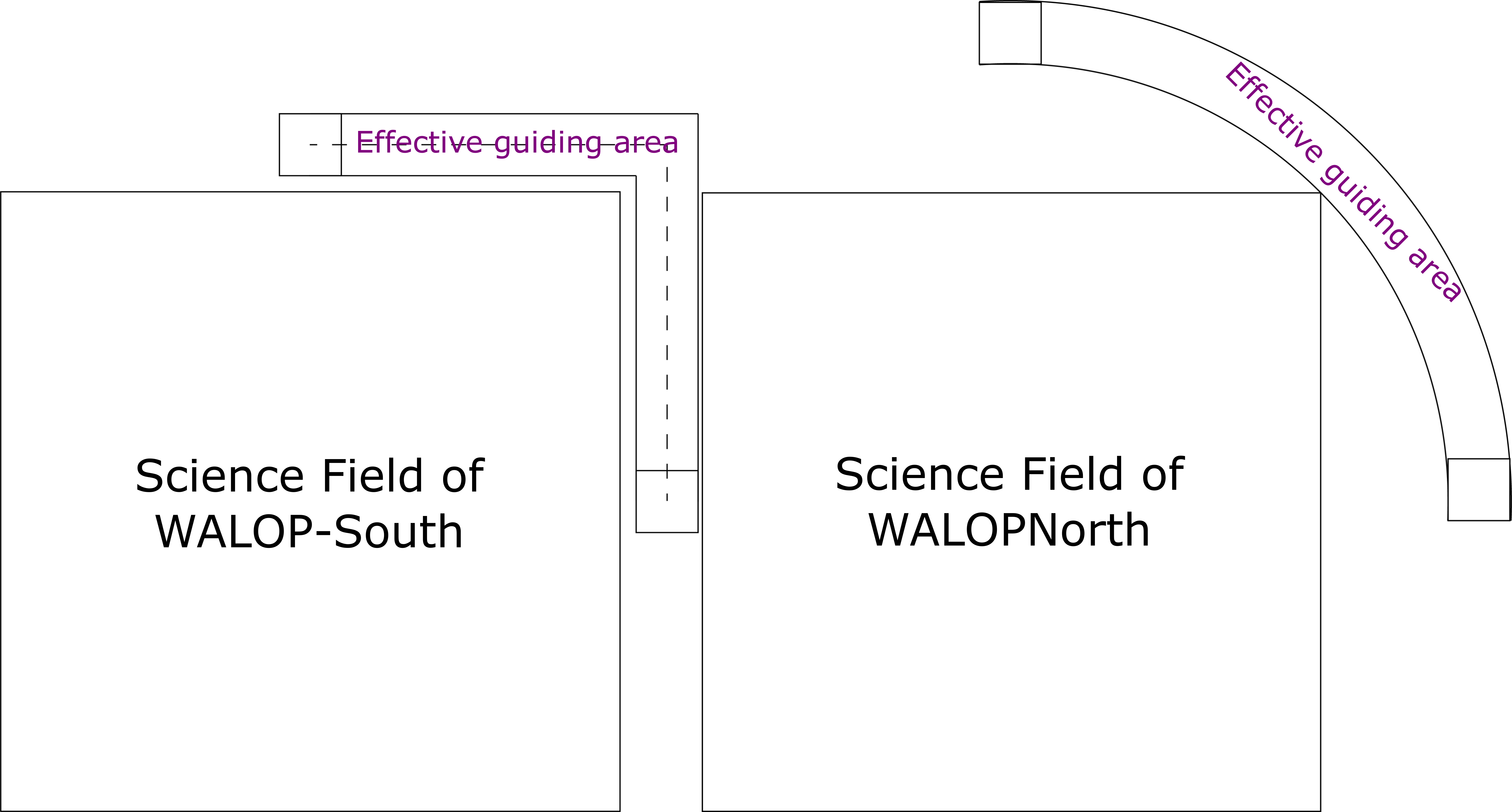}
\end{tabular}
\end{center}
\caption 
{ \label{fig:guidcomp}
Comparison of a trailing guider concept, used in WALOP-South (left) and a rotating guider concept, used in WALOP-North (right). The small squares in each case are the guider fields and the big ones, the science fields in each case (the rotating design's path has been truncated to $90^{\circ{}}$ for illustration purposes).} 
\end{figure}

WALOP-South \cite{SouthOptical} is using an off-axis-pickup guiding system that trails around the science field in a 2-axes XY stage. For WALOP-North, we designed a rotatory system, where the autoguider rotates around the science field. A comparison of the two design approaches is shown in Figure \ref{fig:guidcomp}. Trailing has the advantage of the light-pickup being always as close to the science field as possible and always aligned to the science field. On the other hand, a rotating guider has the advantage of the light pickup being at a set distance to the optical axis at all times (therefore introducing less abberations) and being able to perform a $360^{\circ{}}$ rotation around it (that way increasing the available effective guider FoV).

\section{Instrument performance}\label{sec:perf}

\subsection{PSF Morphology}\label{subsec:psf}
\begin{figure}[!ht]
\begin{center}
\begin{tabular}{c}
\includegraphics[width=0.8\textwidth]{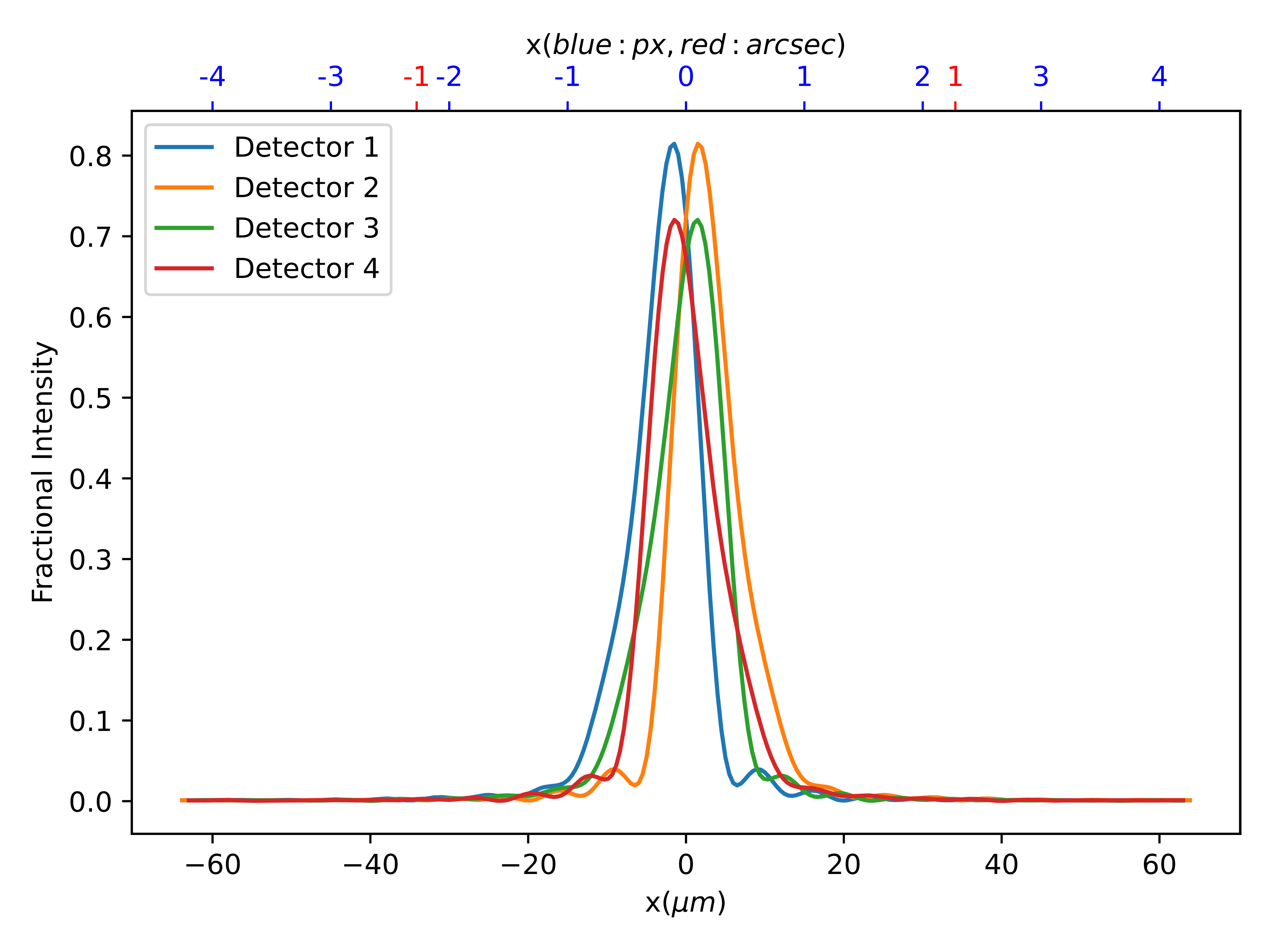}
\end{tabular}
\end{center}
\caption 
{ \label{fig:psf} Plot of the PSF along the x-axis (in $\mu{}m$, $px$, and $arcsec$) for each instrument detector. Each PSF is weight-averaged for wavelength (using the SDSS-r bandpass) and averaged for the entire FoV. Seeing is not included in this calculation. } 
\end{figure}

The first metric of the instrument performance is the PSF width and morphology. The Zemax-generated, wavelength weight-averaged (according to the SDSS-r filter), FoV-averaged PSF of the instrument for all 4 different detectors is shown in Figure \ref{fig:psf}. From this we can see a symmetrical distribution of the instrument PSF, with slight zero-offset at each detector in pairs of 2. Note that this does not take into account the seeing, whose effect will be more apparent in the results of Section \ref{subsec:ensc} .

\begin{figure}[!ht]
\begin{center}
\begin{tabular}{c}
\includegraphics[width=0.8\textwidth]{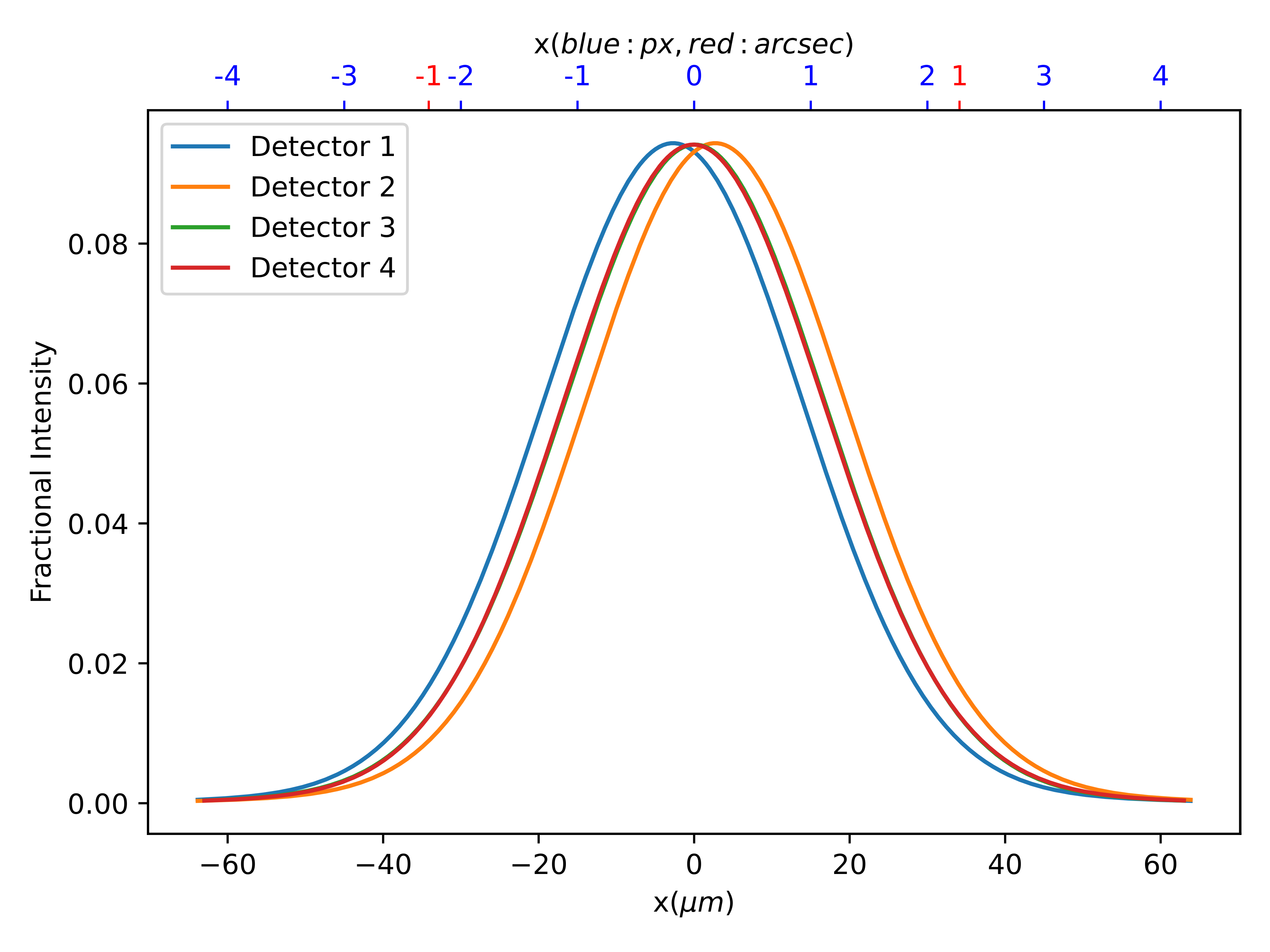}
\end{tabular}
\end{center}
\caption 
{ \label{fig:psf_gauss} \todo{Plot of the PSF along the x-axis (in $\mu{}m$, $px$, and $arcsec$) for each instrument detector. Each PSF is weight-averaged for wavelength (using the SDSS-r bandpass) and averaged for the entire FoV. Seeing is included in this calculation, by means of convolving the seeing-ignorant PSF of Figure \ref{fig:psf} with a Gaussian kernel of FWHM equal to the mean seeing FWHM at Skinakas.}} 
\end{figure}

\todo{In order to account for the seeing in the PSF, we created plots of the instrument PSF convolved with the seeing PSF (used as kernel). We convolved the PSF of Figure \ref{fig:psf} with a Gaussian kernel of $FWHM=1.1''=>\sigma{}=0.4671''$ (producing Figure \ref{fig:psf_gauss}). The FWHM used is the expected median at Skinakas (Table \ref{tab:teldetails}). Comparing Figure \ref{fig:psf_gauss} with Figure \ref{fig:psf} we conclude that our instrument is indeed seeing-limited and that the PSF is accurately sampled as per the requirements\ref{tab:techreq}.}

\todo{The PSF produced by the instrument is well within the requirements of the survey, as it is accurately sampled and seeing-limited. The PSF is also symmetrical and well-behaved across the entire field of view. Nevertheless small positioning errors are apparent in Figure \ref{fig:psf_gauss}. We intend to deal with the differential nature of the PSFs using the instrument's dedicated image analysis software (scope of a follow-up publication), following the data analysis pipeline of the RoboPol instrument \cite{robopipeline}.}

\subsection{Encircled Energy Profiles}\label{subsec:ensc}
\begin{figure}[!ht]%
\centering
\includegraphics[width=0.9\textwidth]{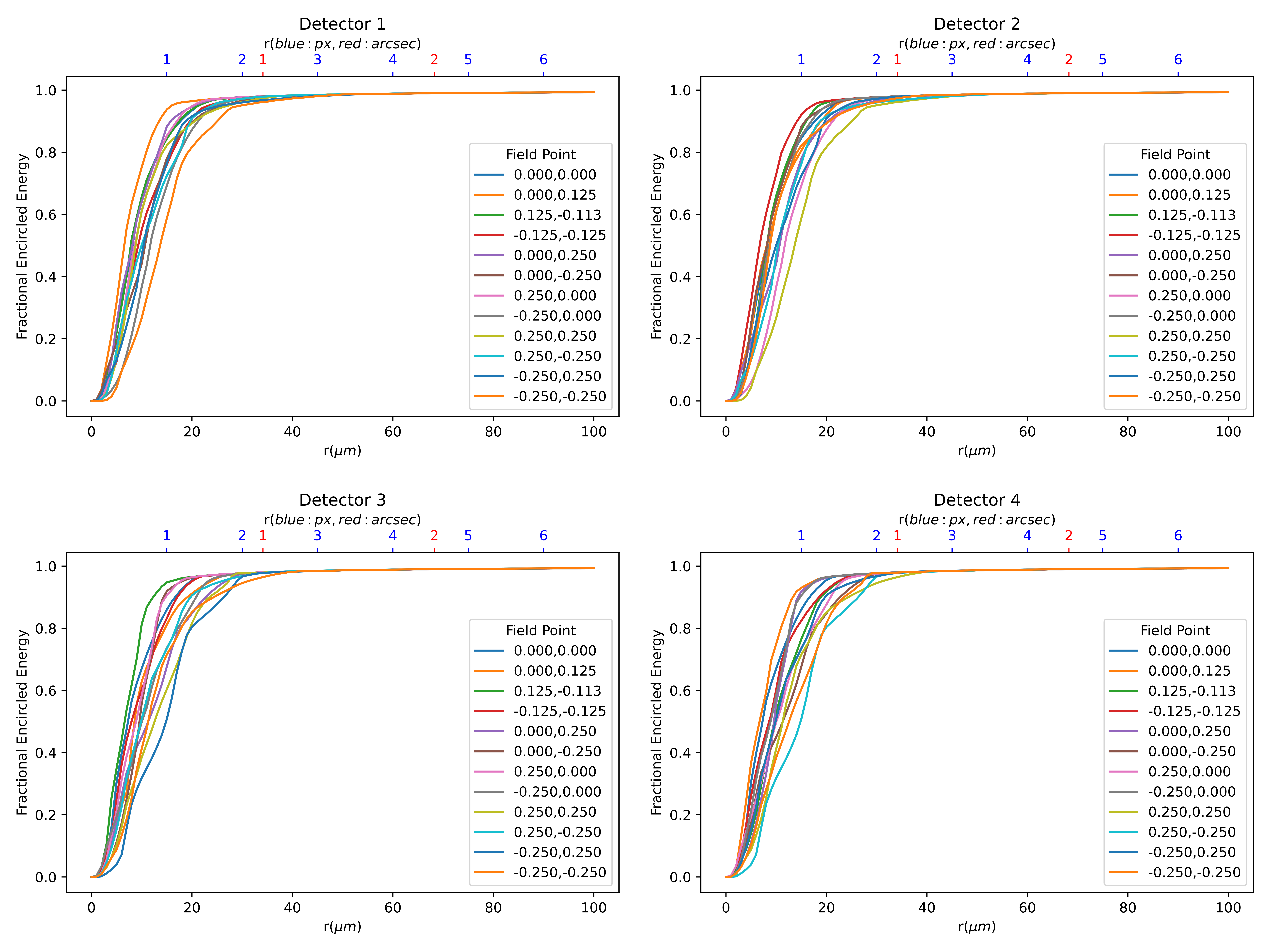}
\caption{Encircled energy profiles for each detector and field position.}
\label{fig:ensc}
\end{figure}

Another measure of the instrument performance is the encircled energy profiles, presented in Figure \ref{fig:ensc}. The encircled energy profile of a spot on a detector gives us the energy detected within a circle, centered at the spot centroid, of radius r with respect to r. These plots were created for the spots produced on each detector, by a simulated star with a Gaussian seeing PSF and FWHM of 1.1'' (as per Section \ref{sec:techreq}). We note that in all cases 99\% of the incident light was recorded within the Nyquist limit (Section \ref{sec:techreq}) and the performance is seeing limited (comparing with the results of Section \ref{subsec:psf}).

\subsection{Spot profiles}\label{subsec:spotprof}
\begin{figure}[!ht]
\begin{center}
\begin{tabular}{c}
\includegraphics[width=0.95\textwidth]{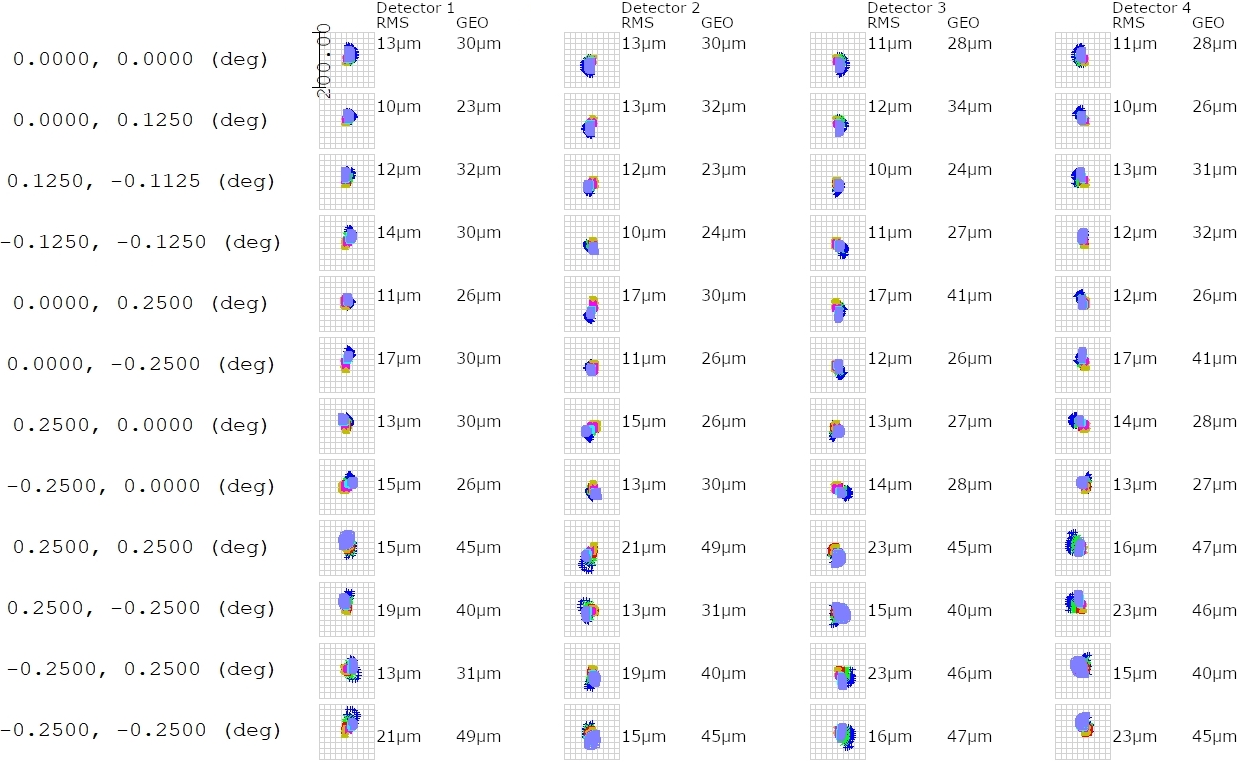}
\end{tabular}
\end{center}
\caption 
{ \label{fig:spot} Spot diagram for the WALOP-North instrument. For every detector (columns) and field point (rows), the spot produced from 10000 simulated rays is depicted. Additionally, the RMS and Geometric radius (GEO) is mentioned next to the spot.} 
\end{figure}

The last metric we will use to assess the optical performance of our instrument is the spot sizes and profiles at different field positions in all 4 detectors we used, presented in Figure \ref{fig:spot}. Most root-mean-square (RMS) radii are below $20\mu{}m$ (1.3px), while most geometric radii are below $40\mu{}m$ (2.7px). We see that the entire field is well behaved and within the limits discussed in Section \ref{sec:techreq}. Note that this calculation was done for a point-source object, similar to Section \ref{subsec:psf} (no seeing effects included).

\subsection{Tolerancing}\label{subsec:toler}
\todo{The instrument performance was also evaluated under various tolerances, originating in misalignments and manufacturing errors. The tolerancing criterion used was the RMS spot size, and the compensator was chosen as the back focus (distance between each CCD and the corresponding final camera lens). The maximum compensation permitted was $1mm$. Table \ref{tab:toler} lists the acceptable tolerance values. A Monte Carlo tolerance analysis with compensation was performed for $10^5$ test systems. Table \ref{tab:tolres} lists the tolerancing results. We see that the criterion was kept below $41\mu{}m$ in all cases. The results show that the instrument is robust against the tolerances considered.}

\begin{table}[ht]
\begin{center}       
\begin{tabular}{|c|c|} 
\hline
\rule[-1ex]{0pt}{3.5ex}  \textbf{Parameter} & \textbf{Tolerancing Value} \\
\hline
\hline
\rule[-1ex]{0pt}{3.5ex}  Radii of Curvature & $\pm{}100\mu{}m$ \\
\hline
\rule[-1ex]{0pt}{3.5ex}  Thicknesses & $\pm{}50\mu{}m$ \\
\hline
\rule[-1ex]{0pt}{3.5ex}  Surface Decentering & $\pm{}50\mu{}m$ \\
\hline
\rule[-1ex]{0pt}{3.5ex}  Element Decentering & $\pm{}30\mu{}m$ \\
\hline
\rule[-1ex]{0pt}{3.5ex}  Surface Tilts & $\pm{}1'$ \\
\hline
\rule[-1ex]{0pt}{3.5ex}  Element Tilts & $\pm{}2'$ \\
\hline
\rule[-1ex]{0pt}{3.5ex}  Surface Irregularities & $1.5$ fringe at $633nm$ \\
\hline
\rule[-1ex]{0pt}{3.5ex}  Indices of Refraction & $\pm{}0.0005$ \\
\hline
\rule[-1ex]{0pt}{3.5ex}  Abbe Numbers & $\pm{}0.5\%$ \\
\hline
\end{tabular}
\caption{\label{tab:toler}\todo{Tolerancing parameters for the WALOP-North instrument.}} 
\end{center}
\end{table}

\begin{table}[ht]
\begin{center}       
\begin{tabular}{|c|c|c|} 
\hline
\rule[-1ex]{0pt}{3.5ex}  \textbf{RMS Radius} & \textbf{Configurations 1\&2} & \textbf{Configurations 3\&4} \\
\hline
\hline
\rule[-1ex]{0pt}{3.5ex}  Nominal & $10.30\mu{}m$ & $10.18\mu{}m$ \\
\hline
\rule[-1ex]{0pt}{3.5ex}  Best & $10.43\mu{}m$ & $10.44\mu{}m$ \\
\hline
\rule[-1ex]{0pt}{3.5ex}  Worst & $40.88\mu{}m$ & $36.04\mu{}m$ \\
\hline
\rule[-1ex]{0pt}{3.5ex}  Mean & $17.87\mu{}m$ & $17.45\mu{}m$ \\
\hline
\rule[-1ex]{0pt}{3.5ex}  Standard Deviation & $3.58\mu{}m$ & $3.29\mu{}m$ \\
\hline
\end{tabular}
\caption{\label{tab:tolres}\todo{Results of the Monte Carlo tolerance analysis for the WALOP-North instrument. The criterion used was the RMS spot size. The maximum compensation of the back focus was $1mm$.}} 
\end{center}
\end{table}

\section{WALOP Filters}\label{sec:filt}
\subsection{Choice of Filter}\label{subsec:filterchoice}
The SDSS-r filter has been selected for use with both WALOP instruments. Johnson-Cousins filters have very broad profiles (compared to the SDSS ones). Given the wavelength-dependent dispersion of birefringent optics, a wider filter would make the design substantially more difficult despite being superior in terms of transmitted light. We determined the PSF-integrated photon flux for the SDSS-r and Johnson-Cousins-R filters for a survey model star to securely exclude the Johnson-Cousins-R filter. This was selected to be a SDSS-r $16.5 mag$ (cut-off magnitude for the survey), $5180K$ (mean temperature of stars in PASIPHAE's intended area, according to the Besançon model for stellar population synthesis of the Galaxy\cite{bec}) star. The PSFs used were those produced by Zemax for the optimal instrument design for each filter\cite{SouthOptical}. At the Skinakas observatory, we also calculated the value for the sky photon flux\cite{skisky}. With those estimations, we could determine the mean (over the entire field) photometric signal-to-noise ratio (SNR) for the instrument in either filter for a 20-minute exposure. The results are reported in Table \ref{tab:filterphot}. Given its superior PSF characteristics and apparent performance parity with the Johnson-Cousins-R filter, the SDSS-r filter was selected for WALOPs.

\begin{table}[ht]
\begin{center}       
\begin{tabular}{|c|c|c|} 
\hline
\rule[-1ex]{0pt}{3.5ex}  & \textbf{Cousins-R} & \textbf{SDSS-r}  \\
\hline
\hline
\rule[-1ex]{0pt}{3.5ex}  \textbf{Star Flux} $\left(\frac{ph}{s\cdot{}cm^2}\right)$ & 0.278 & 0.225  \\
\hline
\rule[-1ex]{0pt}{3.5ex}  \textbf{Sky Flux} $\left(\frac{ph}{s\cdot{}cm^2\cdot{}arcsec^2}\right)$ & 0.00779 & 0.00560  \\
\hline
\rule[-1ex]{0pt}{3.5ex}  \textbf{SNR} & 168 & 149  \\
\hline
\end{tabular}
\caption{\label{tab:filterphot}Photometric Calculations for the WALOP candidate filters.} 
\end{center}
\end{table}

\subsection{Making the filter non-polarizing}\label{subsubsec:nonpolsdss}
\begin{figure}[!ht]
\begin{center}
\begin{tabular}{c}
\includegraphics[width=0.8\textwidth]{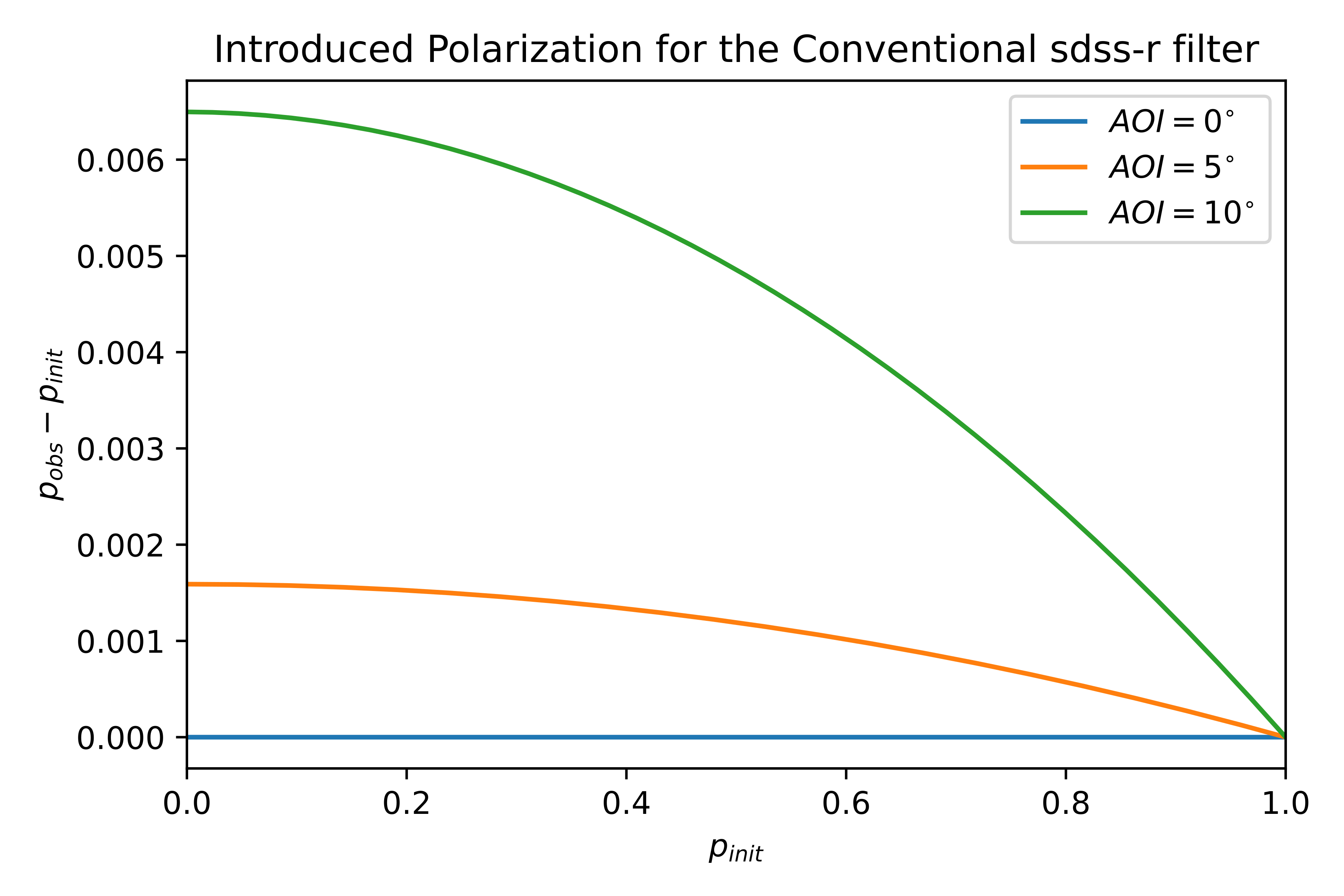}
\end{tabular}
\end{center}
\caption 
{ \label{fig:convsdssintrod}
The polarization introduced by the conventional SDSS-r filter in different AOI as a function of the input polarization. $p_{init}$ is the incident polarization to the filter and $p_{obs}$ is the observed polarization after the filter (by an ideal polarimeter).} 
\end{figure}

Glass substrates are covered with thin coatings to provide commercially available SDSS filters. This impacts off-axis beam polarization in proportion to the incidence angle. The filter is positioned inside the collimator for WALOP-North, making it subject to beams with different incidence angles. Therefore, we had to ensure that the filter we use would not cause the polarization to increase past the instrument's $p=0.1\%$ systematics goal. Given the transmittance curve of the filter at various incidence angles, we can calculate the polarization introduced by the filter at any incident angle for any incident polarization fraction. The highest polarization introduced by a commercial SDSS-r filter (at a $10^\circ{}$ angle of incidence (AOI), which is the extreme angle of incidence on the WALOP filters, as per the design) was found to be above $0.6\%$ (Figure \ref{fig:convsdssintrod}). This does not comply with the instrument's limit; hence the creation of a modified SDSS-r filter was required.

\begin{figure}[!ht]
\begin{center}
\begin{tabular}{c}
\includegraphics[width=0.8\textwidth]{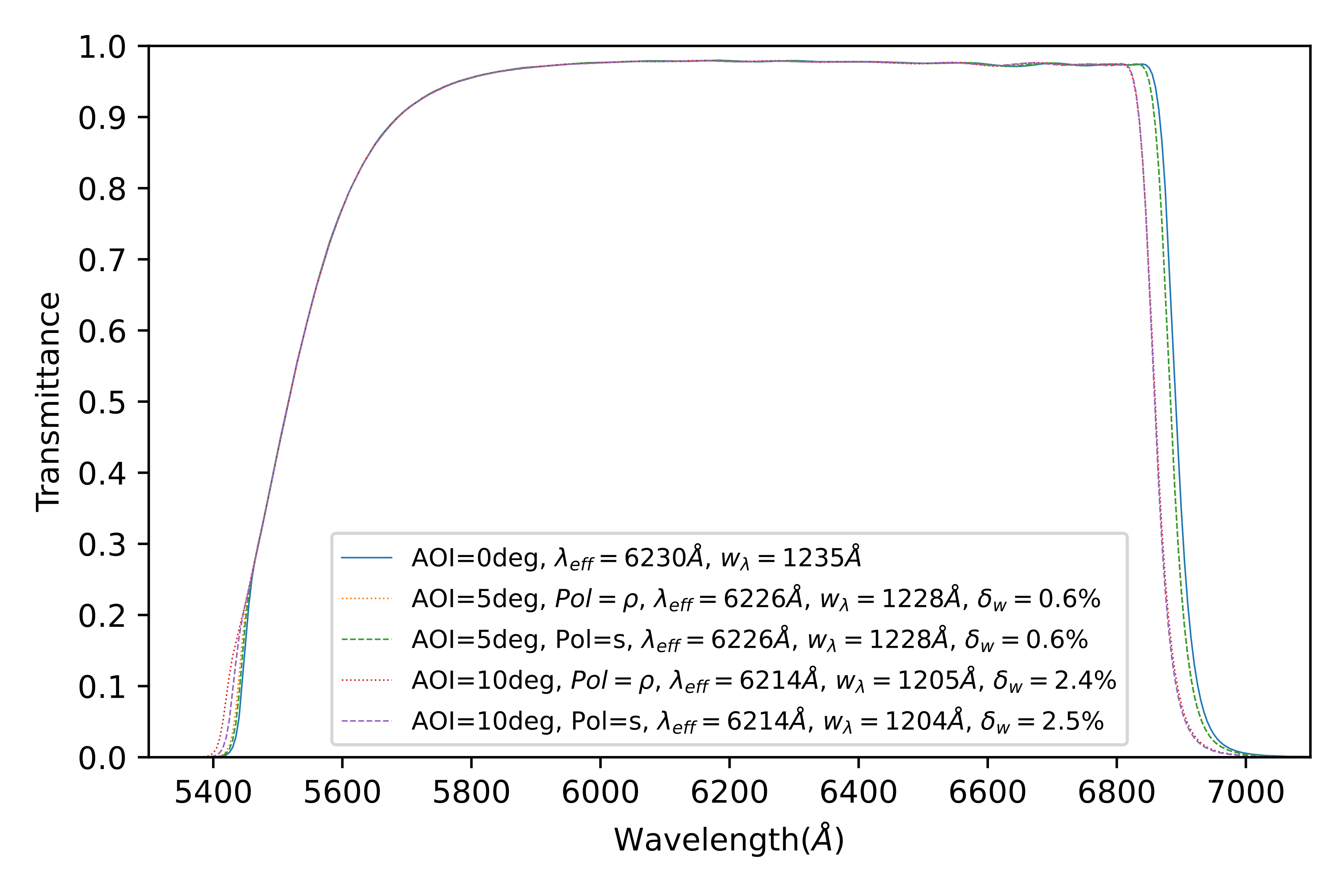}
\end{tabular}
\end{center}
\caption 
{ \label{fig:npsdsstrans}
\todo{The transmittance curve of the new non-polarizing SDSS-r filter in two perpendicular polarization states ($\rho{}$ and $s$) and angles of incidence (lines intentionally thin to help distinguish the graphs). $\lambda{}_{eff}$ is the effective wavelength of the filter, $w_{\lambda{}}$ is the filter wavelength-width, and $\delta{}_{w}$ is the wavelength-width difference between each respective $AOI$ and $AOI=0^{\circ{}}$. Note: we chose the nomenclature "$\rho$" and "$s$" for the filter's perpendicular polarization states instead of the conventional "$o$" and "$e$" to reflect the difference of reference frame between the filter and Wollaston prism (since both these components have a separate polarization angle reference frame). The reference frame of the Wollaston prism corresponds with the EVPA reference frame.}} 
\end{figure}

\begin{figure}[!ht]
\begin{center}
\begin{tabular}{c}
\includegraphics[width=0.8\textwidth]{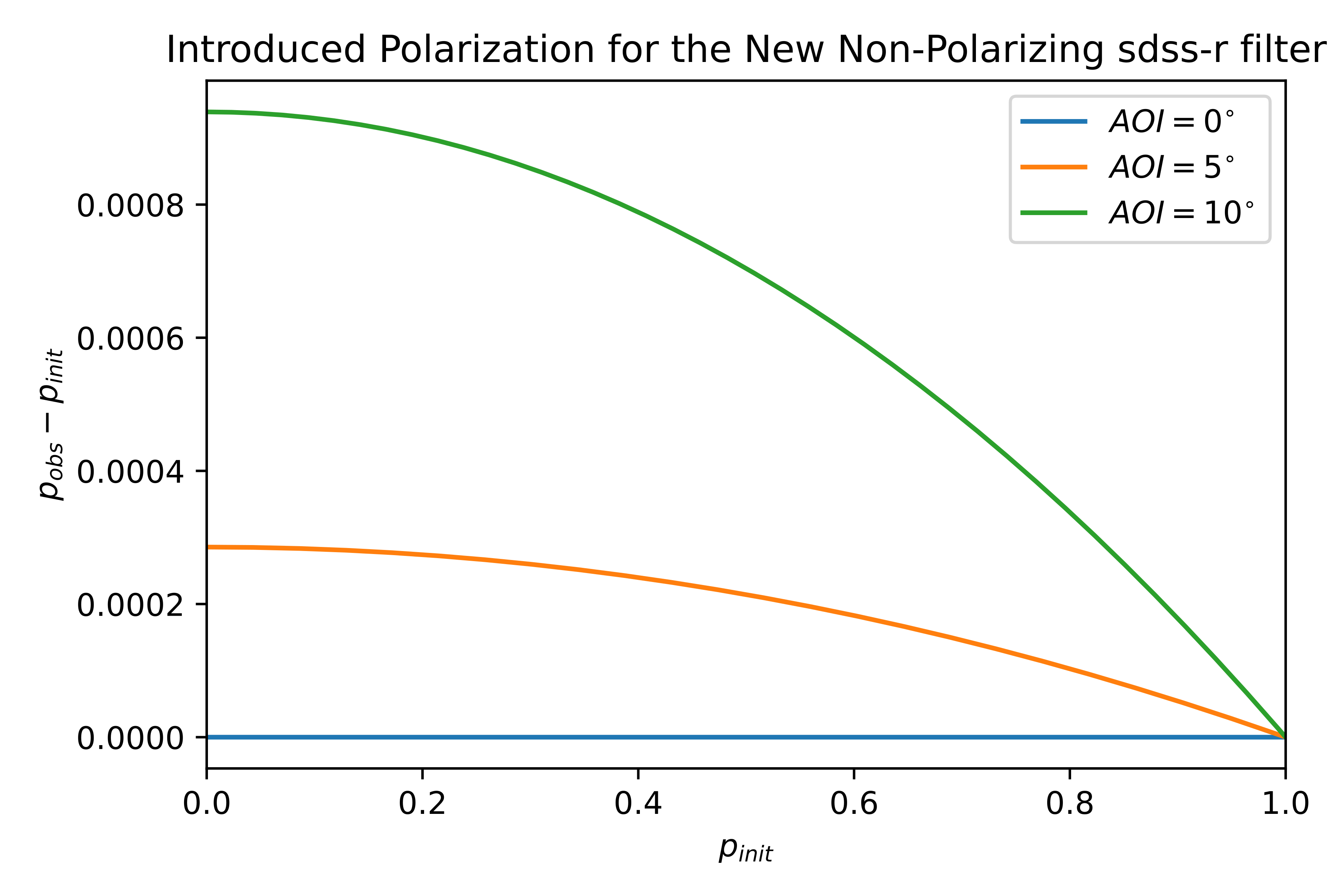}
\end{tabular}
\end{center}
\caption 
{ \label{fig:npsdssintrod}
The polarization introduced by the new non-polarizing SDSS-r filter in different angles of incidence as a function of the input polarization.} 
\end{figure}

We were able to design an SDSS-r filter with much lower induced polarization together with Asahi Spectra Company\cite{asahi}. The design of that filter consisted of fine-tuning the dielectric coatings so that as low as possible polarization is introduced, while retaining a profile as similar as possible to the original SDSS-r. Figures \ref{fig:npsdsstrans} and \ref{fig:npsdssintrod} demonstrate its transmittance curve (which is identical to the usual SDSS-r filter) and added polarization respectively.

\subsection{Manufactured Filters}\label{subsec:realfilt}
For our filters additional requirements must be met, since they have to perform optically in addition to their spectral performance. They must not distort the image or have any optical power, like a lens would. For the needs of the PASIPHAE project, 3 such filters were commissioned to Asahi Spectra. One for each instrument (hereby filters A1 and A2) and a backup filter (hereby filter B1). Appendix \ref{app:filt} shows the performance of these filters after manufacture. A summary of the interferometry and spectroscopy results is presented in Table \ref{tab:interfspec}. We see that filters A1 and A2 are very similar and up to par with the instrument requirements. Filter B1 is performing a bit worse, yet also within the required limits for the instrument.

\begin{table}[ht]
\begin{center}       
\begin{tabular}{|c|c|c|c|} 
\hline
\rule[-1ex]{0pt}{3.5ex} \textbf{Filter}  & \textbf{Max Optical Power} & \textbf{Max RMS Wavefront} & \textbf{max$\left(p_{obs}-p_{init}\right)$} \\
\hline
\hline
\rule[-1ex]{0pt}{3.5ex}  \textbf{A1} & 0.114 waves & 0.0035 waves & 0.025\%  \\
\hline
\rule[-1ex]{0pt}{3.5ex}  \textbf{A2} & 0.077 waves & 0.026 waves & 0.025\%  \\
\hline
\rule[-1ex]{0pt}{3.5ex}  \textbf{B1} & 0.128 waves & 0.038 waves & 0.030\% \\
\hline
\end{tabular}
\caption{\label{tab:interfspec}Interferometric and spectroscopic performance of the manufactured WALOP filters.} 
\end{center}
\end{table}

\section{Polarimetric Calibration}\label{sec:calib}
\todo{In order to calibrate the instrument, we will follow a similar methodology as prescribed for WALOP-South\cite{WALOP_Calibration_paper,birefr}. This is a numerical approach and not an analytical Stokes-\todos{Mueller}\cite{soleillet} one, due to the instrument complexity. The following sections present the methodology we applied for the calibration of WALOP-North.}

\subsection{Zemax Simulations}\label{subsec:calibzemax}
Zemax OpticStudio provides the tools to simulate the system's \todo{(instrument+telescope)} performance. We can use OpticStudio\footnote{in our case version 13} to calculate the output intensity in one of the 4 CCD detectors, as a function of input intensity in a specific set of 12 field points per iteration. The drawbacks of simulating the instrument using OpticStudio are the following:
\begin{itemize}
    \item Each iteration will provide information on the output light intensity for 1 detector
    \item A limited number of field points (12) are available per iteration
    \item The light of the input beam is limited to being either unpolarized or fully polarized at a set polarization angle
\end{itemize}

\todo{In order to overcome those limitations, we ran multiple sequential simulations each on different polarization state, field points and detectors. In the end, we combined the data from all detectors and field points, and interpolated the data from different polarizations (as shown in section \ref{subsec:calibeq}) to extract the calibration of our instrument. That way, we can gain information for polarization states between 0 and 100\% polarized, as well as infer the calibration for all field points between the limited amount simulated.}

\subsection{Zemax Inputs}\label{subsec:zmxinp}
\begin{figure}[!ht]
\begin{center}
\begin{tabular}{c}
\includegraphics[width=0.8\textwidth]{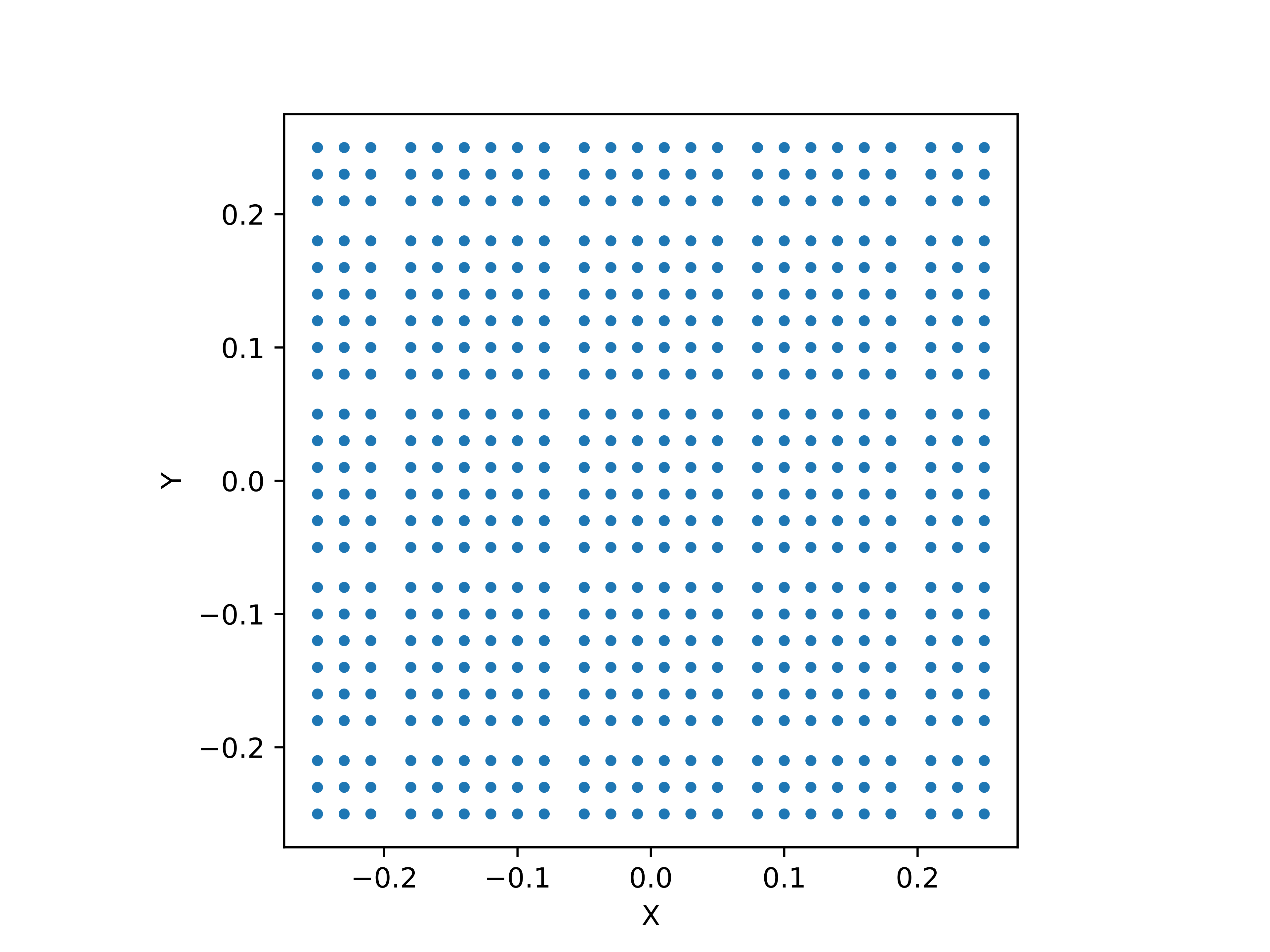}
\end{tabular}
\end{center}
\caption 
{ \label{fig:infield}
The FoV sampling we input to Zemax. (units of field are degrees)} 
\end{figure}

\begin{figure}[!ht]
\begin{center}
\begin{tabular}{c}
\includegraphics[width=0.8\textwidth]{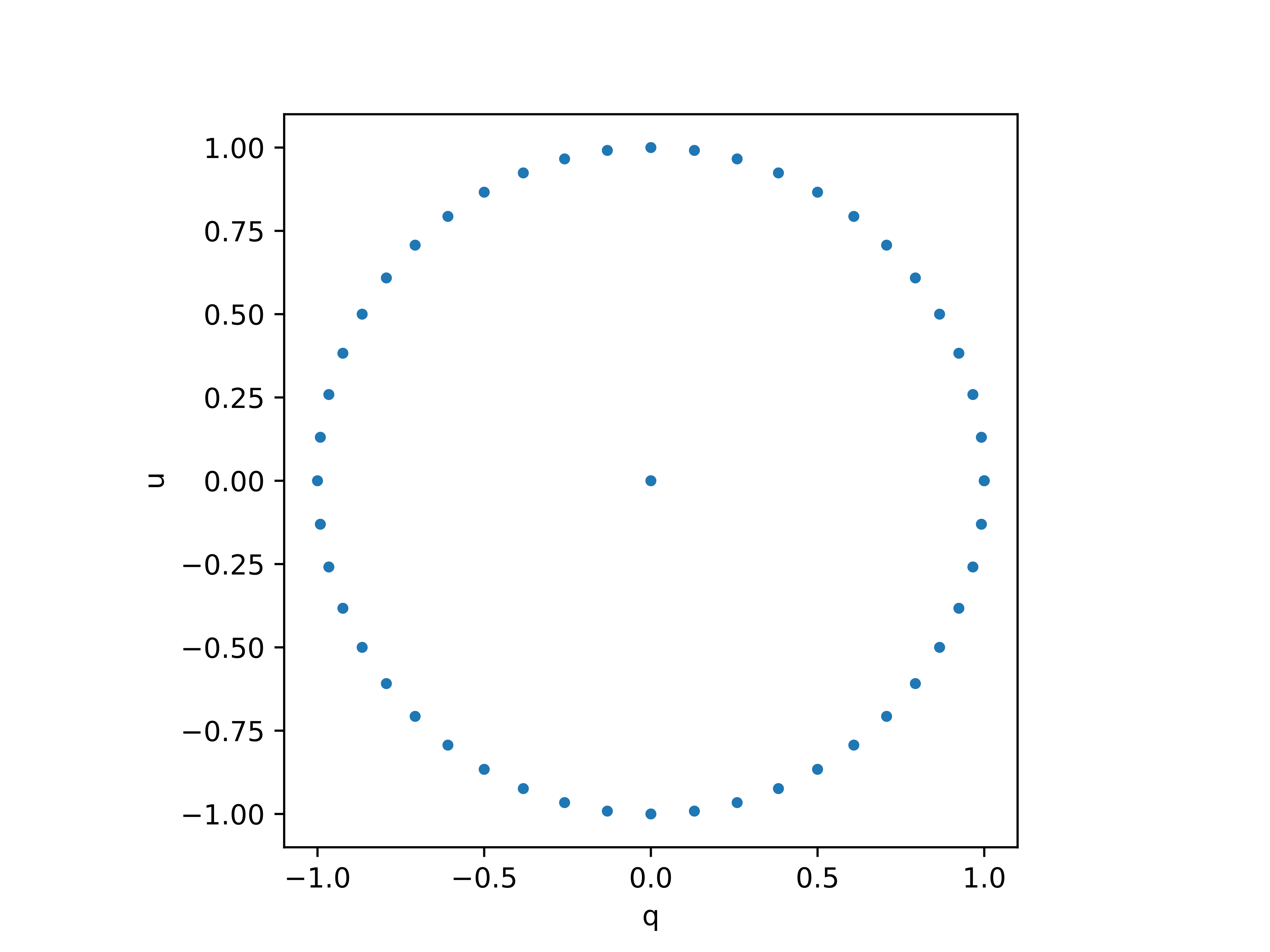}
\end{tabular}
\end{center}
\caption 
{ \label{fig:inpol}
The polarization sampling we input to Zemax.} 
\end{figure}

\todo{The first step to running our tests was to generate the inputs for Zemax. We created the Zemax input files, containing the spatial coordinates of the field points sampling the entire FoV, and the polarization states simulated for every field position (all either 100\% or 0\% polarized). Then, we compiled a ZPL (Zemax Programming Language) script that took care of recursively feeding all the polarization and field coordinates to Zemax and producing Transmittance files for each detector. These files give us the total instrument transmittance at different wavelengths (within the SDSS-r' band, with effective wavelength dictated by the model star and the filter's band-pass) for each combination of detectors (4 total), field points (576 total - 1.25' sampling of the entire FoV) and polarization (50 total - 49 polarized states and 1 unpolarized) provided. We therefore ran a total of 115200 simulations, each of $10^9$ rays. Figure \ref{fig:infield} depicts the FoV sampling and Figure \ref{fig:inpol} depicts the polarization sampling of the simulations.}

\subsection{Calibration Equations}\label{subsec:calibeq}
We have now acquired the transmittance at each detector ($D$ ranging from $1$ to $4$: integer), each field point ($x$, $y$ ranging from -0.25 to 0.25 degrees) and each input fully polarized (or unpolarized) state ($q$, $u$ ranging $-1$ to $1$ such that $\sqrt{q^2+u^2}$= $0$ or $1$). We shall name this transmittance $T_{D,x,y,q,u}$. We denote the collection of all measured transmittances with a common parameter by replacing this parameter with '*' (grouping all consecutive *s), e.g. the collection of all measured transmittances of detector $2$, will be named $T_{2,*}$, while all measured transmittances of detector $2$, at the center point will be $T_{2,0,0,*}$.

We denote $q_r$, $u_r$ the input \todos{normalized} Stokes parameters to the Zemax simulation. We define as $q_o$, $u_o$ the \todos{normalized} Stokes parameters as observed by the uncalibrated instrument. For every field point and every input polarization, the observed parameters are calculated as:
\begin{equation}\label{eq:obsq}
    q_o=\frac{T_{3,*}-T_{1,*}}{T_{3,*}+T_{1,*}}
\end{equation}
\begin{equation}\label{eq:obsu}
    u_o=\frac{T_{2,*}-T_{4,*}}{T_{2,*}+T_{4,*}}
\end{equation}

\todos{These differ from $q_r$ and $u_r$, due to various instrument systematics caused by the dependence of polarization behavior of optical components on factors like angle of incidence and propagation of the beam, wavelength, birefringence variation across the beam etc. \cite{WALOP_Calibration_paper,SouthOptical,birefr}. As it is impractical to analytically model the combined effect of all this from first principles, we fit} the input $q$ and $u$ parameters as second degree polynomials of $q_o$ and $u_o$ (using the Levenberg-Marquardt algorithm\cite{lm1,lm2} through Scipy\cite{scipy}):
\begin{equation}\label{eq:qmodel}
    q_r = a_0+a_qq_o+a_uu_o+a_{q^2}q^2_o+a_{u^2}u^2_o+a_{qu}q_ou_o
\end{equation}
\begin{equation}\label{eq:umodel}
    u_r = b_0+b_qq_o+b_uu_o+b_{q^2}q^2_o+b_{u^2}u^2_o+b_{qu}q_ou_o
\end{equation}

The goal is to recover the coefficients $a$ and $b$ that will allow the mapping of the observed Stokes parameters to the input ones. To this model fitting, we add a $0.1\%{}$ noise (by sampling $q_o$ and $u_o$ from a Gaussian with $0.1\%$ spread), as this is the level of expected noise to be faced during on-sky calibration. In contrast to WALOP-South's calibration\cite{WALOP_Calibration_paper}, the mixed term $\left(qu\right)$ has been added to the model. This is to reduce the fit's covariance.

\subsection{Verification of  Calibration}\label{subsec:calibverif}
In order to verify statistically our calibration, we need to be able to feed the model with mock observed data and compare the modeled against the theoretical instrument input $q_i$ and $u_i$. Furthermore, the instrument input needs to be partially polarized, much like the real cases the instrument will face.

To create this mock data, we first fit the transmittances discussed in Section \ref{subsec:calibeq} as second degree polynomials of $q_r$ and $u_r$:
\begin{equation}\label{eq:t1}
    T_{1,*,q_r,u_r} = a^{T}_{1,0}+a^{T}_{1,q}q_r+a^{T}_{1,u}u_r+a^{T}_{1,q^2}q^2_r+a^{T}_{1,u^2}u^2_r+a^{T}_{1,qu}q_ru_r
\end{equation}
\begin{equation}\label{eq:t2}
    T_{2,*,q_r,u_r} = a^{T}_{2,0}+a^{T}_{2,q}q_r+a^{T}_{2,u}u_r+a^{T}_{2,q^2}q^2_r+a^{T}_{2,u^2}u^2_r+a^{T}_{2,qu}q_ru_r
\end{equation}
\begin{equation}\label{eq:t3}
    T_{3,*,q_r,u_r} = a^{T}_{3,0}+a^{T}_{3,q}q_r+a^{T}_{3,u}u_r+a^{T}_{3,q^2}q^2_r+a^{T}_{3,u^2}u^2_r+a^{T}_{3,qu}q_ru_r
\end{equation}
\begin{equation}\label{eq:t4}
    T_{4,*,q_r,u_r} = a^{T}_{4,0}+a^{T}_{4,q}q_r+a^{T}_{4,u}u_r+a^{T}_{4,q^2}q^2_r+a^{T}_{4,u^2}u^2_r+a^{T}_{4,qu}q_ru_r
\end{equation}

We then generate a mock set of $q_i$ and $u_i$, sampling the entire Stokes plane (including partial polarization states that could not be included in the set of $q_r$, $u_r$). We feed these parameters to the transmittance model (Equations \ref{eq:t1}-\ref{eq:t4}), in place of $q_r$ and $u_r$, generating a set of modeled transmittances: $T^m_{1,*}$-$T^m_{4,*}$. From them, we calculate the mock observed $q$ and $u$ ($q^m_o$ and $u^m_o$), based on Equations \ref{eq:obsq} and \ref{eq:obsu}, replacing $T_{1,*}$-$T_{4,*}$. These are then fed into the model (Equations \ref{eq:qmodel} and \ref{eq:umodel}), replacing $q_o$ and $u_o$ to produce the calibrated polarization ($q_c$ and $u_c$), which is then compared to the mock input polarization ($q_i$ and $u_i$).

\subsection{Calibration Results}\label{subsec:calibres}
\begin{figure}[!ht]%
\centering
\includegraphics[width=0.9\textwidth]{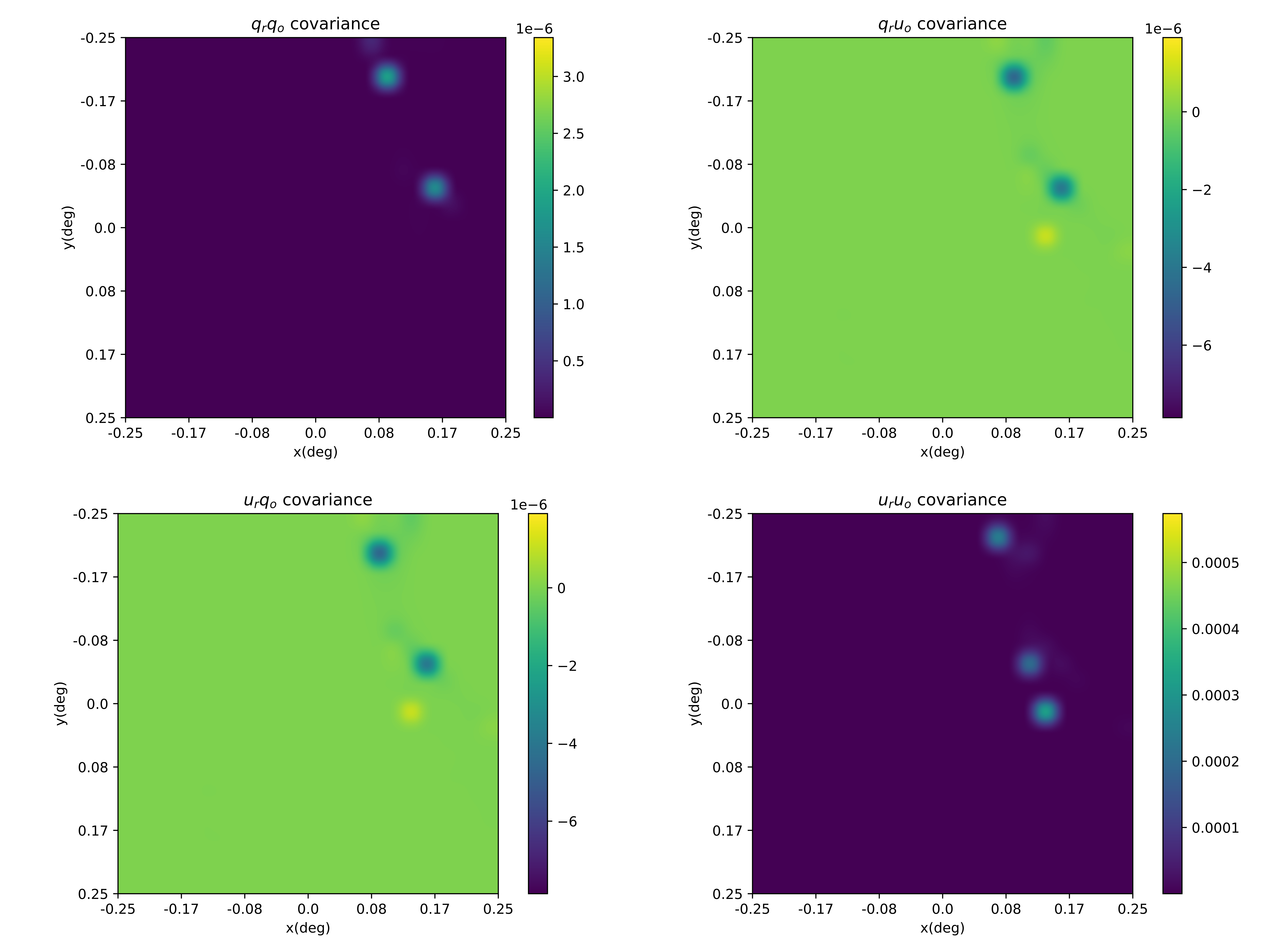}
\caption{Maps of the elements of the covariance matrix from the fit on Equations \ref{eq:qmodel} and \ref{eq:umodel}.}
\label{fig:covar}
\end{figure}

The first measure we used to evaluate our calibration model, is the covariance of the fit of Equations \ref{eq:qmodel} and \ref{eq:umodel}, created by the fit itself described in Section \ref{subsec:calibeq}. The covariance matrix is presented in Figure \ref{fig:covar}. From that, we understand that the fit is very well behaved, as no large covariance exists over the entire FoV. Some spots near the top right corner have a slightly larger covariance, especially in their $qu$ term. This hints to a higher instrumental cross-talk in this region (discussed later in this Section). The non-vanishing covariance presented is the reason we chose to transition to a model that includes the $qu$ term, as opposed to the calibration strategy of WALOP-South\cite{WALOP_Calibration_paper}.

\begin{figure}[!ht]%
\centering
\includegraphics[width=0.9\textwidth]{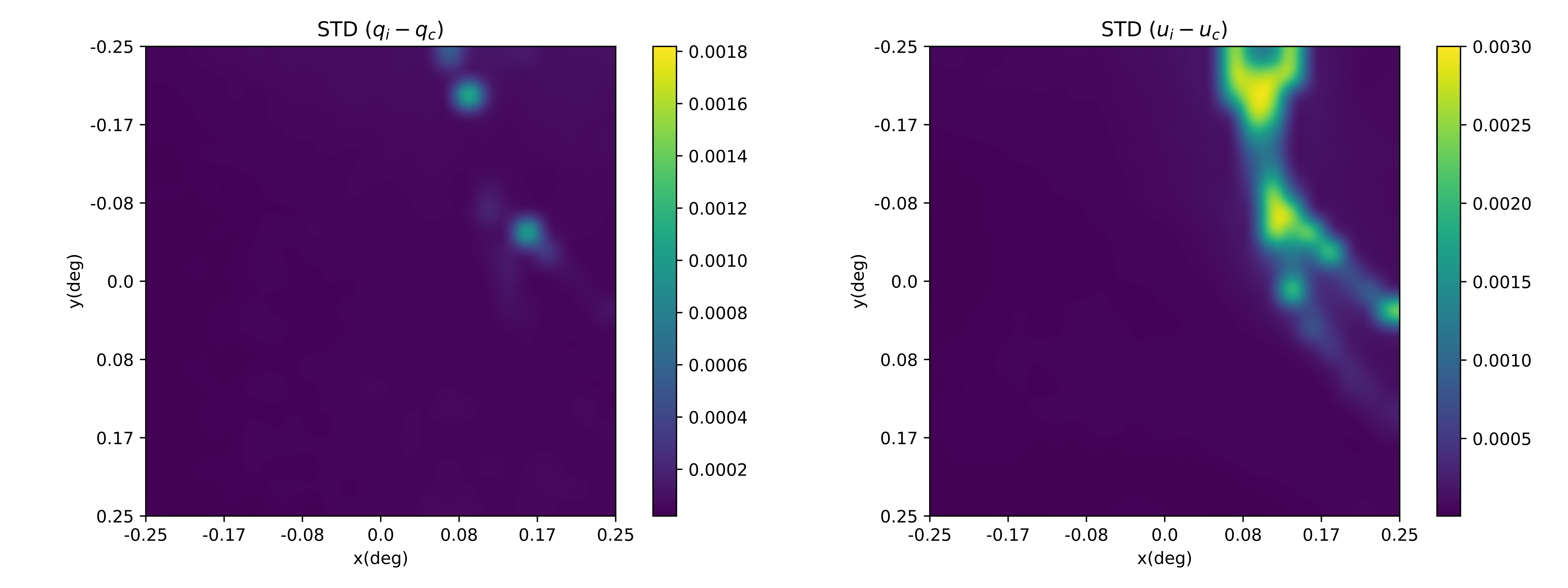}
\caption{Maps of the standard deviations of the difference between the input and the retrieved $q$ and $u$ ($a$ and $b$ respectively) from the calibration model. The closer to 0, the better.}
\label{fig:std}
\end{figure}

The second metric of the success of the calibration is the standard deviation of the difference between the input $\left(q_i,u_i\right)$ and the retrieved $\left(q_c,u_c\right)$ parameters, for varying degrees of input polarization as described in Section \ref{subsec:calibverif}. This is depicted in Figure \ref{fig:std}. From that, we understand that the retrieved parameters adhere well to the inputs and therefore the instrument can be calibrated using the described strategy. There is a problematic area over an arc between the top and right edge of the FoV. This region's problematic behaviour is due to the large cross-talk between $u$ and $q$ at the half-wave plate (HWP) of the instrument's polarizing assembly at this region.

\begin{figure}[!ht]%
\centering
\includegraphics[width=0.9\textwidth]{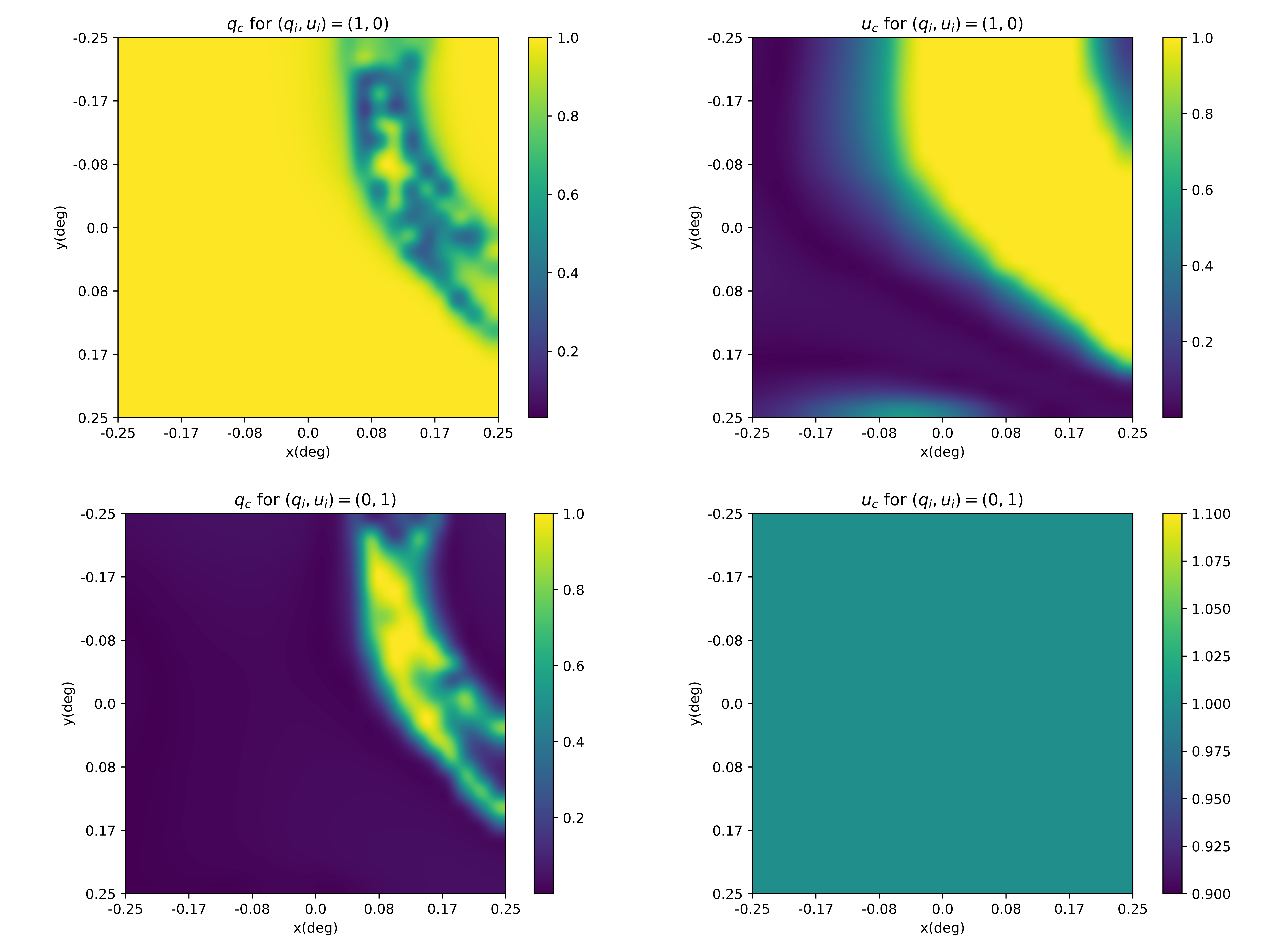}
\caption{Maps of the retrieved parameters for the 4 polar fully polarized input parameters. Maps (a) and (d) depict the throughput of the instrument, while (b) and (c) the cross-talk.}
\label{fig:cross}
\end{figure}

The final measure we use are the maps of the retrieved parameters for the following 2 fully polarized input parameters: $\left(q_i,u_i\right)=\left(1,0\right)$ and $\left(0,1\right)$. These are depicted in Figure \ref{fig:cross}. The maps of $q_c$ and $u_c$ for $\left(q_i,u_i\right)=\left(1,0\right)$ and $\left(q_i,u_i\right)=\left(0,1\right)$ respectively are essentially the instrument's throughput (the amount of polarization in a specified Stokes parameter that gets recorded by the instrument, without interference from the other parameter). The maps of $q_c$ and $u_c$ for $\left(q_i,u_i\right)=\left(0,1\right)$ and $\left(q_i,u_i\right)=\left(1,0\right)$ respectively are essentially the instrument's cross-talk (the amount of polarization in a specified Stokes parameter that gets recorded by the instrument, only as result of the other parameter's value). From that, we understand that indeed the problematic areas in the calibration are due to very high crosstalk at the region in the shape of an arc between the top and right edge of the FoV. This is similar to the result for WALOP-South\cite{WALOP_Calibration_paper}, which utilizes the same Wollaston prism assembly as the currently discussed instrument.

\subsection{On-Sky Calibration}\label{subsec:skycal}
\todo{After commissioning, the instrument will be able to follow the same polarimetric calibration procedure (as will WALOP-South\cite{WALOP_Calibration_paper}), using a polarizer mounted at the instrument entrance, before all other optics.}

\todo{This polarizer will be able to convert light from a dense stellar field (which sufficiently samples the FoV), irrespective of the polarization of stars, to fully-polarized. In conjunction with the polarizer's rotation, this creates a set of fully polarized inputs at any Electric Vector Polarization Angle (EVPA).}

\todo{These observations, combined with observations of stars with known polarization\cite{robostandards} and of polarimetric flat field sources such as the bright sky around near full-Moon creates a collection of observations of input sources with various polarizations. As part of developing the calibration protocol for WALOP-South, we have shown that the sky around the Full Moon within a range of two days can serve as a highly effective polarimetric flat source (with an accuracy greater than 0.05\% in p) for wide-field polarimeter calibration \cite{moonlight}. While the polarimetric sky flats allow for relative calibration of the full FoV, observing standard stars across the FoV allows absolute calibration for the full FoV.}

\todo{Following the paradigm described in the current Section, we will be able to calibrate the instrument within the $0.1\% STD$ instrumental level required, as shown in Figure \ref{fig:std}. Furthermore, we have verifed this calibration methodolgy in a lab-prototype of WALOP with accuracies  matching the calibration model prediction\cite{WALOP_Calibration_paper}.}

\section{Conclusion}
We have presented the optical design of WALOP-North, a wide-field linear optical polarimeter with a field of view $0.25\times{}0.25$ $deg^2$, capable of reaching measured polarization accuracy of $0.1\%$. The polarimeter will be used by the PASIPHAE collaboration to carry out a polarimetric survey of the Galaxy. The instrument had design challenges, as it is intended to fit challenging spatial, polarimetric, and imaging constrains, which we have shown it is up to par with.

\todo{We showed a new design for filters that adhere to the SDSS-r standard, while introducing minimal polarization to the wide field. The optical performance of the filters is also acceptable, since they introduce no significant aberration to the wavefront. Their design and testing methods we developed are extensible to other filters, as necessary for the polarimetric needs of WALOP-North and other polarimeters.}

\todo{We also detailed a calibration strategy that accomplishes the strict requirements of the instrument performance, with minimal instrumental polarization in the end-result. We calibrate the predicted instrument data to the desired accuracy. The procedure is suitable for on sky calibration of the instrument. Finally, the procedure is also applicable to the general-case scenario of calibrating large-FoV polarimeters.}

\appendix    

\section{Manufactured Filter Performance}\label{app:filt}
\subsection{A1 Filter}\label{app:a1filt}
\begin{figure}[!ht]
\begin{center}
\begin{tabular}{c}
\includegraphics[width=0.8\textwidth]{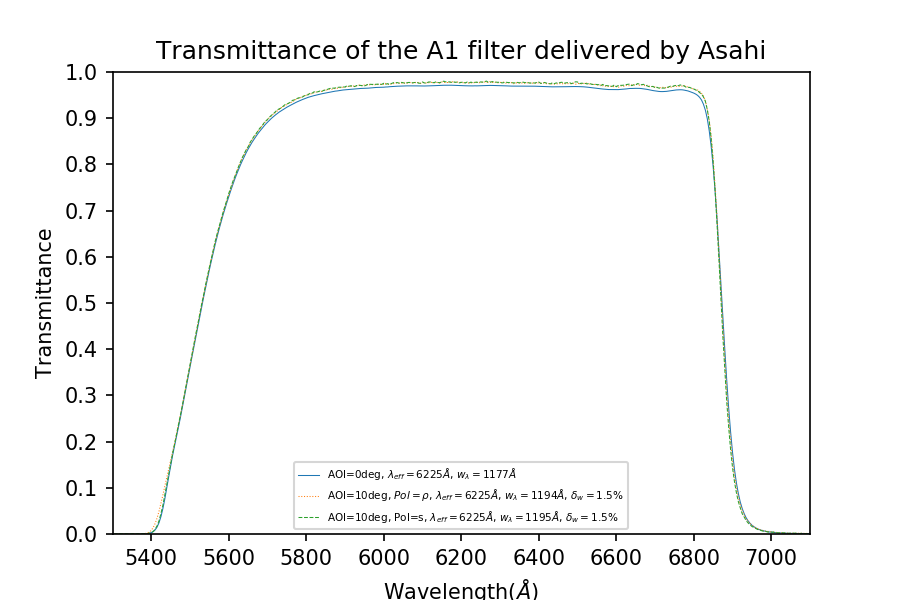}
\end{tabular}
\end{center}
\caption 
{ \label{fig:a1trans}
The transmittance curve of the A1 filter in different polarization states and 2 different angles of incidence.} 
\end{figure}

\begin{figure}[!ht]
\begin{center}
\begin{tabular}{c}
\includegraphics[width=0.8\textwidth]{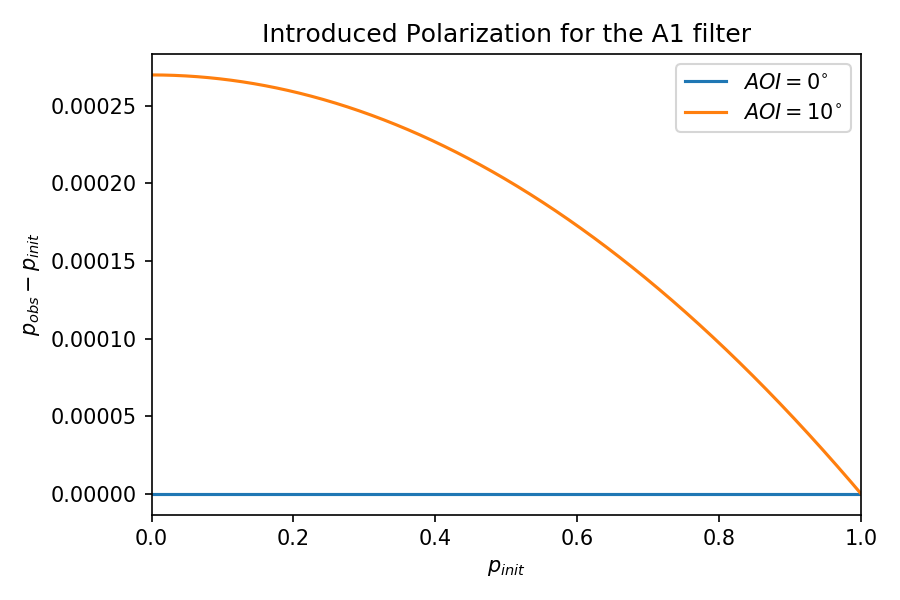}
\end{tabular}
\end{center}
\caption 
{ \label{fig:a1introd}
The polarization introduced by the A1 filter in 2 different angles of incidence (normal and extreme) as a function of the input polarization.} 
\end{figure}

The transmittance curve and introduced polarization of the A1 filter are shown in Figures \ref{fig:a1trans} and \ref{fig:a1introd} respectively.

\begin{figure}[!ht]
\begin{center}
\begin{tabular}{c}
\includegraphics[width=0.7\textwidth]{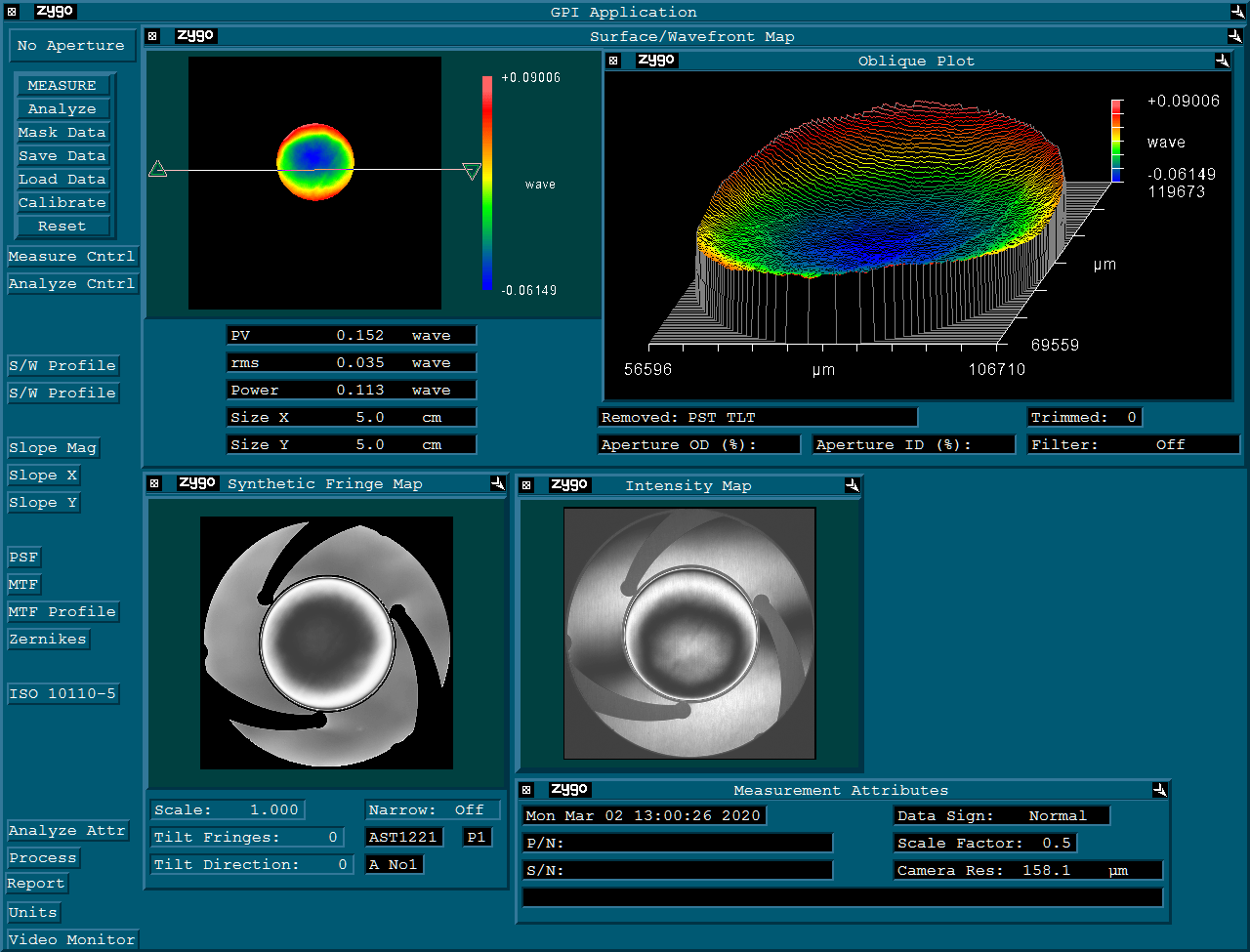}
\end{tabular}
\end{center}
\caption 
{ \label{fig:a1interf1}
The first interferometric and wavefront measurement of the A1 filter.} 
\end{figure}

\begin{figure}[!ht]
\begin{center}
\begin{tabular}{c}
\includegraphics[width=0.7\textwidth]{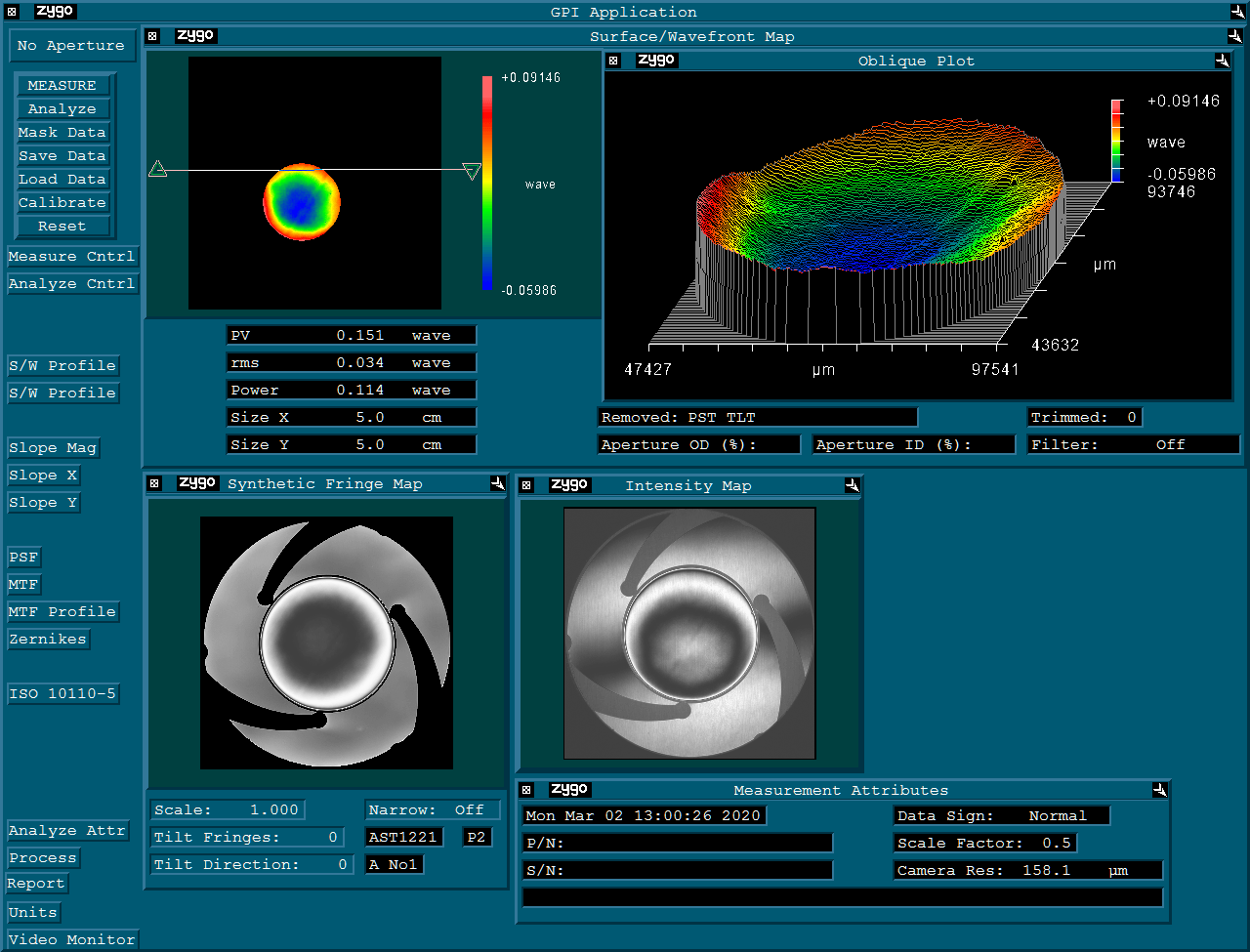}
\end{tabular}
\end{center}
\caption 
{ \label{fig:a1interf2}
The second interferometric and wavefront measurement of the A1 filter.} 
\end{figure}

\begin{figure}[!ht]
\begin{center}
\begin{tabular}{c}
\includegraphics[width=0.7\textwidth]{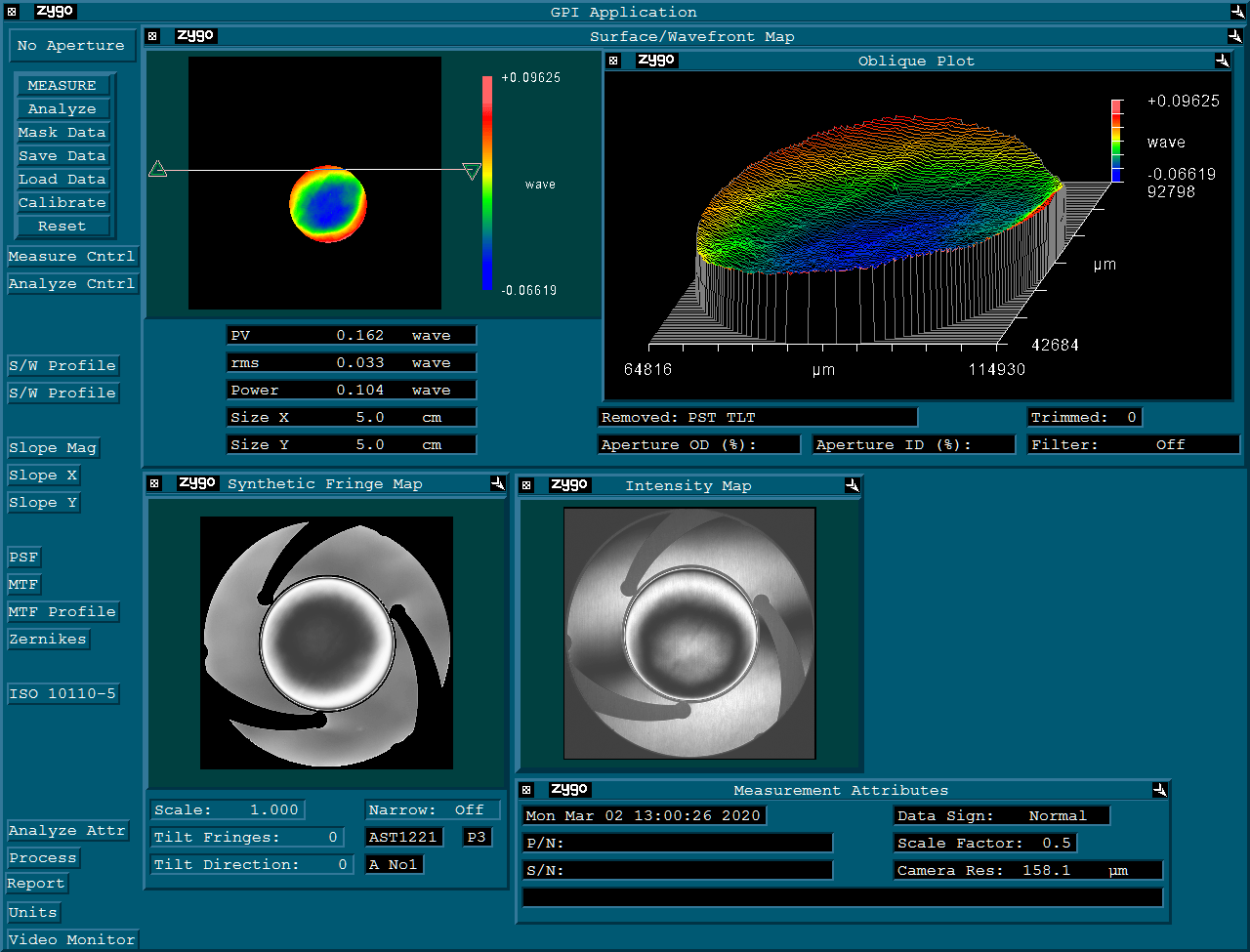}
\end{tabular}
\end{center}
\caption 
{ \label{fig:a1interf3}
The third interferometric and wavefront measurement of the A1 filter.} 
\end{figure}

The optical interferometry\cite{interferometry} at 3 different locations and wavefront measurement of the A1 filter are shown in Figures \ref{fig:a1interf1}, \ref{fig:a1interf2} and \ref{fig:a1interf3}.

\subsection{A2 Filter}\label{app:a2filt}
\begin{figure}[!ht]
\begin{center}
\begin{tabular}{c}
\includegraphics[width=0.8\textwidth]{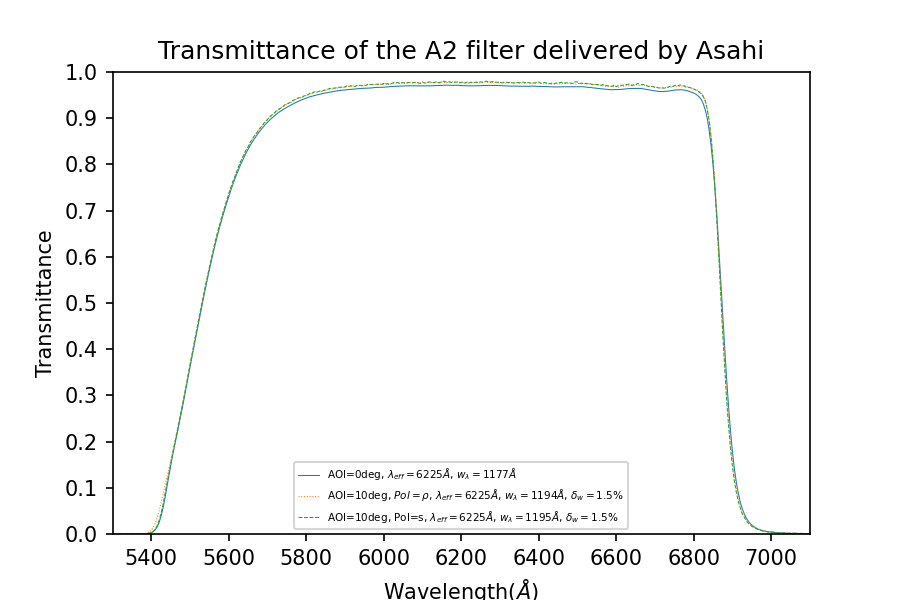}
\end{tabular}
\end{center}
\caption 
{ \label{fig:a2trans}
The transmittance curve of the A2 filter in different polarization states and 2 different angles of incidence.} 
\end{figure}

\begin{figure}[!ht]
\begin{center}
\begin{tabular}{c}
\includegraphics[width=0.8\textwidth]{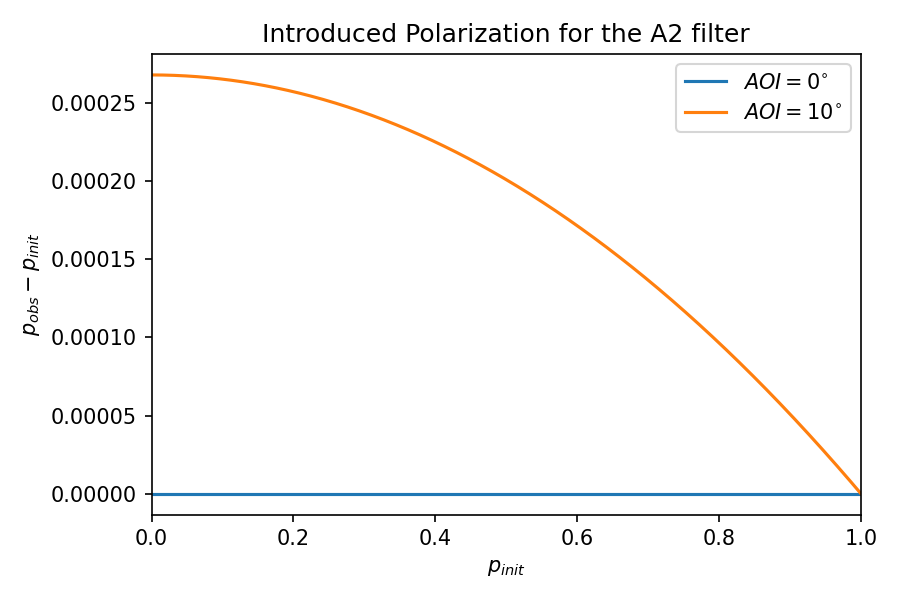}
\end{tabular}
\end{center}
\caption 
{ \label{fig:a2introd}
The polarization introduced by the A2 filter in 2 different angles of incidence (normal and extreme) as a function of the input polarization.} 
\end{figure}

The transmittance curve and introduced polarization of the A2 filter are shown in Figures \ref{fig:a2trans} and \ref{fig:a2introd} respectively.

\begin{figure}[!ht]
\begin{center}
\begin{tabular}{c}
\includegraphics[width=0.7\textwidth]{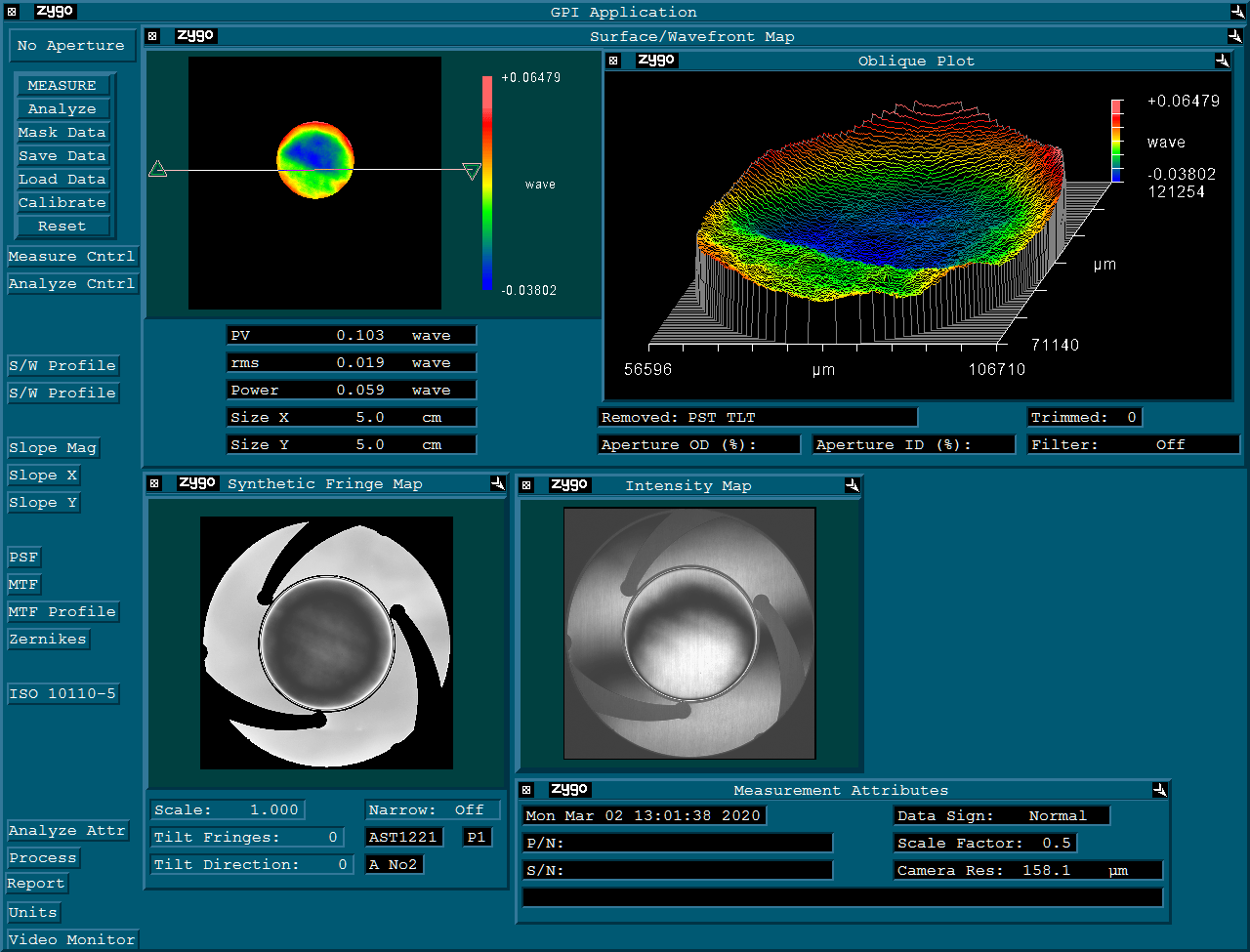}
\end{tabular}
\end{center}
\caption 
{ \label{fig:a2interf1}
The first interferometric and wavefront measurement of the A2 filter.} 
\end{figure}

\begin{figure}[!ht]
\begin{center}
\begin{tabular}{c}
\includegraphics[width=0.7\textwidth]{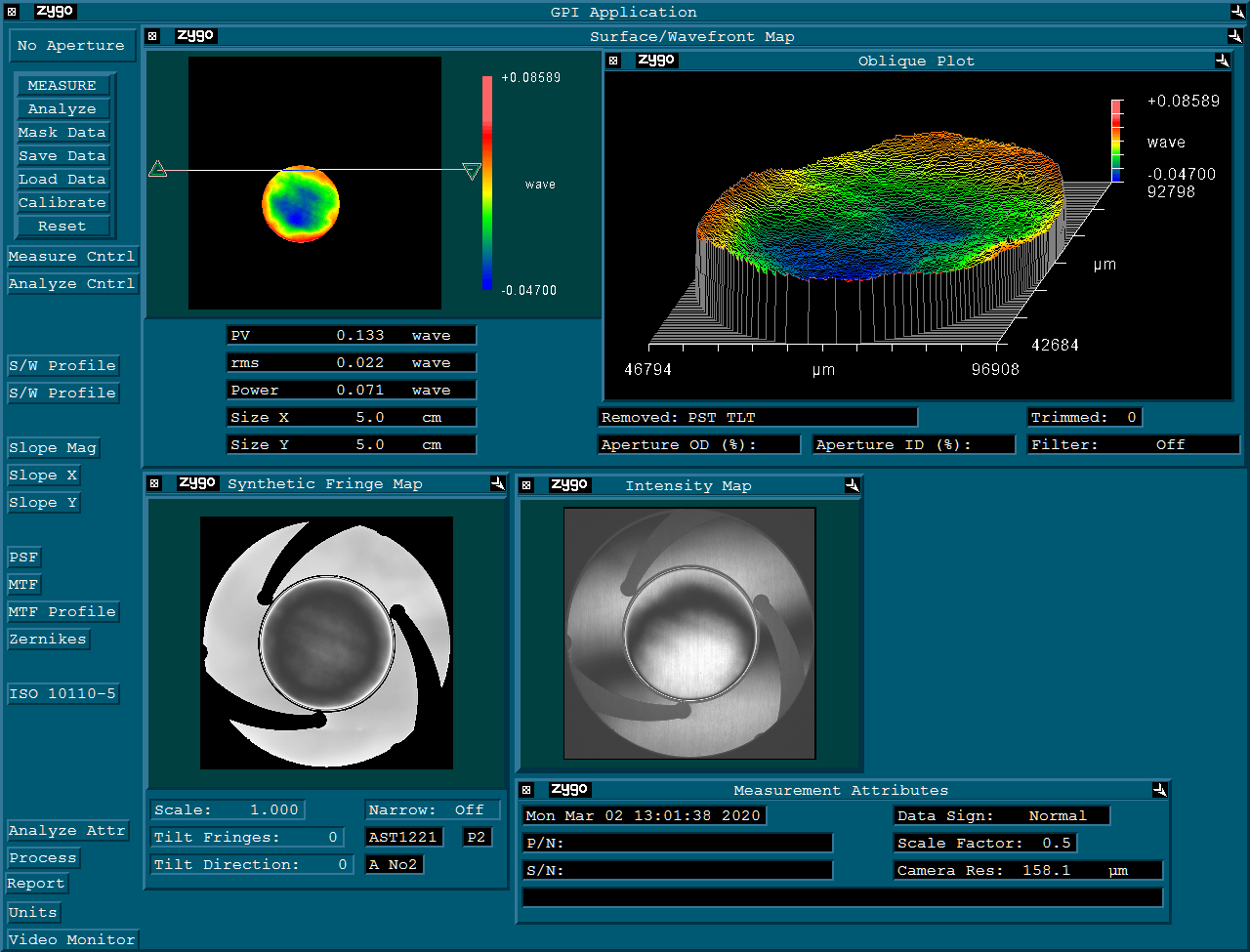}
\end{tabular}
\end{center}
\caption 
{ \label{fig:a2interf2}
The second interferometric and wavefront measurement of the A2 filter.} 
\end{figure}

\begin{figure}[!ht]
\begin{center}
\begin{tabular}{c}
\includegraphics[width=0.7\textwidth]{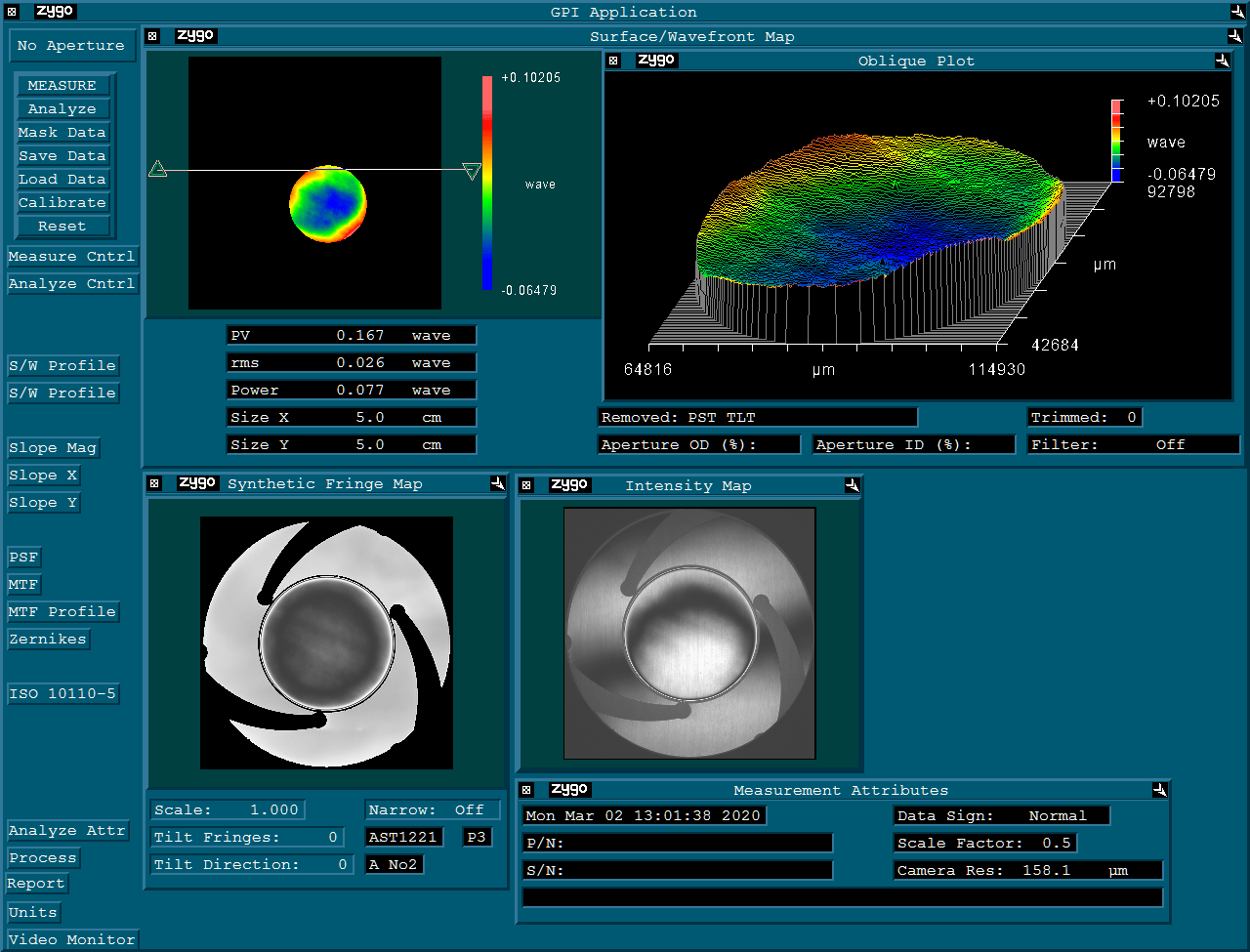}
\end{tabular}
\end{center}
\caption 
{ \label{fig:a2interf3}
The third interferometric and wavefront measurement of the A2 filter.} 
\end{figure}

The 3-level interferometry and wavefront measurement of the A2 filter is shown in Figures \ref{fig:a2interf1}, \ref{fig:a2interf2} and \ref{fig:a2interf3}.

\subsection{B1 Filter}\label{app:b1filt}
\begin{figure}[!ht]
\begin{center}
\begin{tabular}{c}
\includegraphics[width=0.7\textwidth]{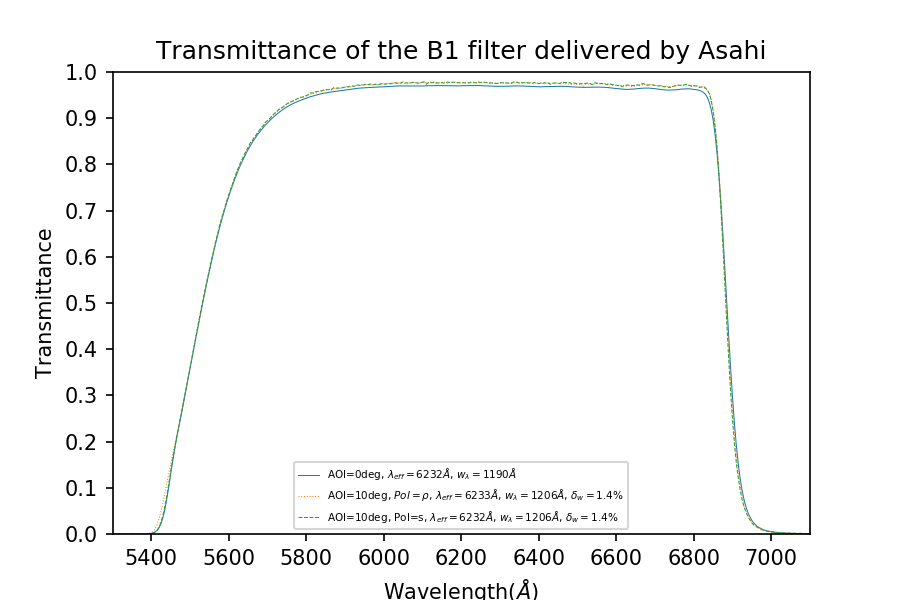}
\end{tabular}
\end{center}
\caption 
{ \label{fig:b1trans}
The transmittance curve of the B1 filter in different polarization states and 2 different angles of incidence.} 
\end{figure}

\begin{figure}[!ht]
\begin{center}
\begin{tabular}{c}
\includegraphics[width=0.7\textwidth]{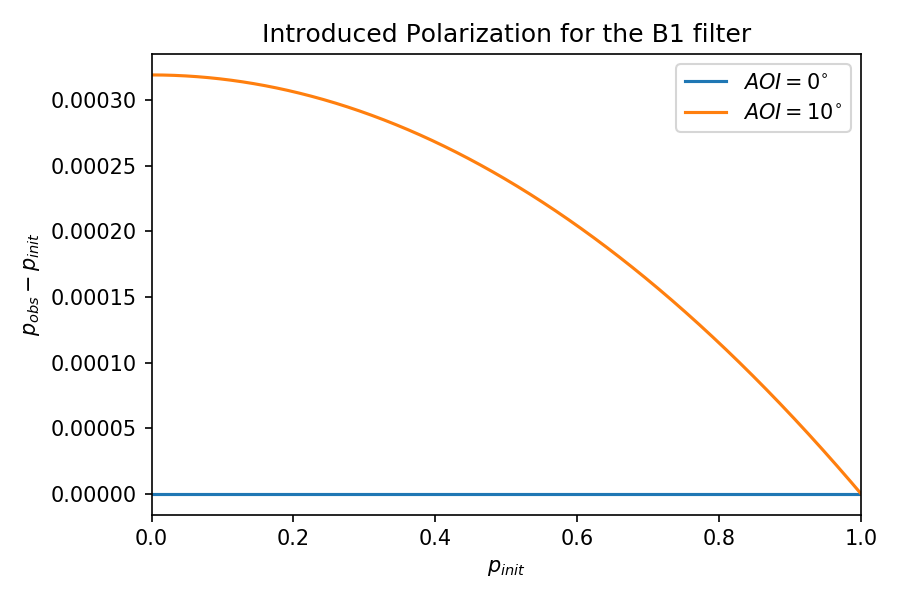}
\end{tabular}
\end{center}
\caption 
{ \label{fig:b1introd}
The polarization introduced by the B1 filter in 2 different angles of incidence (normal and extreme) as a function of the input polarization.} 
\end{figure}

The transmittance curve and introduced polarization of the B1 filter are shown in Figures \ref{fig:b1trans} and \ref{fig:b1introd} respectively.

\begin{figure}[!ht]
\begin{center}
\begin{tabular}{c}
\includegraphics[width=0.7\textwidth]{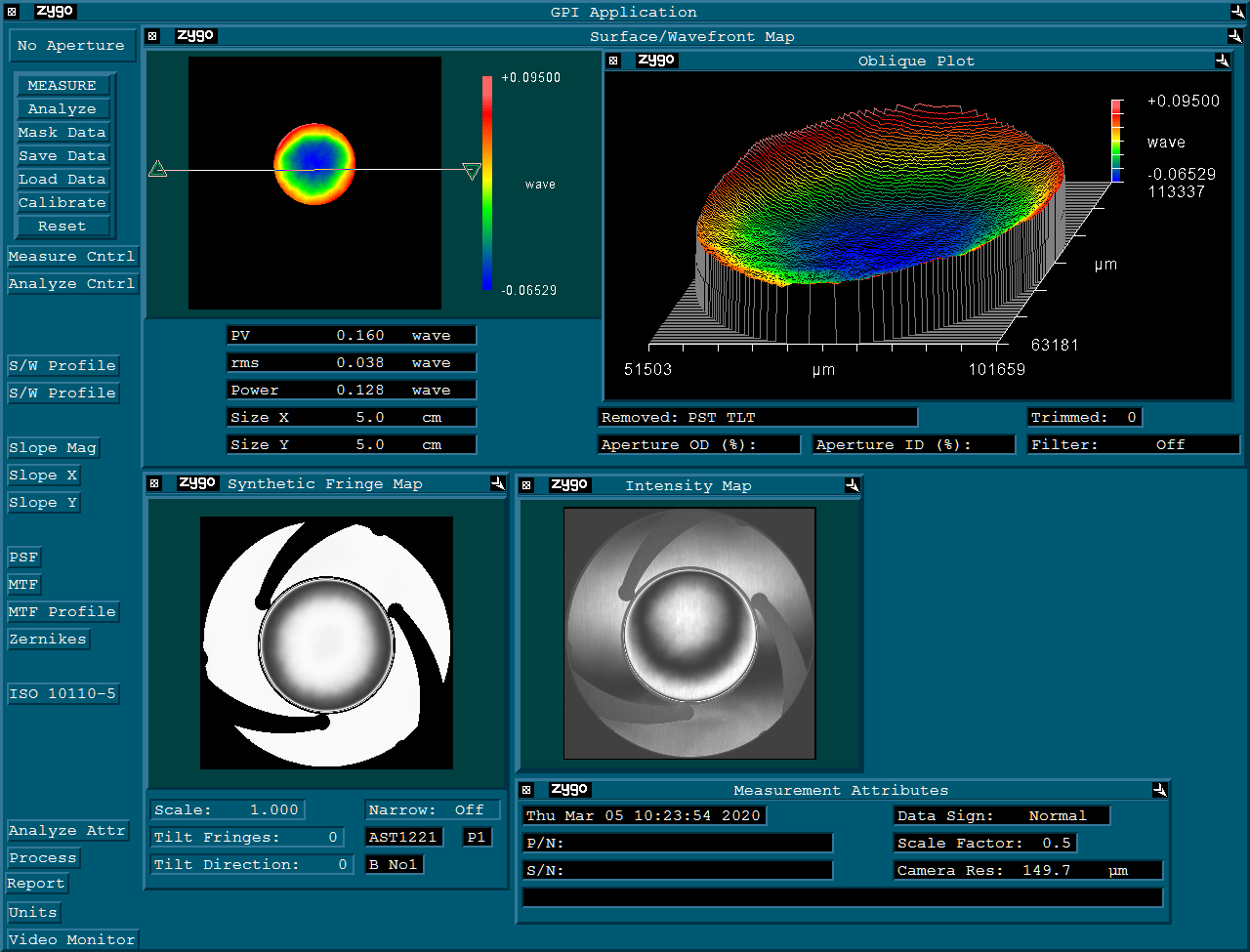}
\end{tabular}
\end{center}
\caption 
{ \label{fig:b1interf1}
The first interferometric and wavefront measurement of the B1 filter.} 
\end{figure}

\begin{figure}[!ht]
\begin{center}
\begin{tabular}{c}
\includegraphics[width=0.7\textwidth]{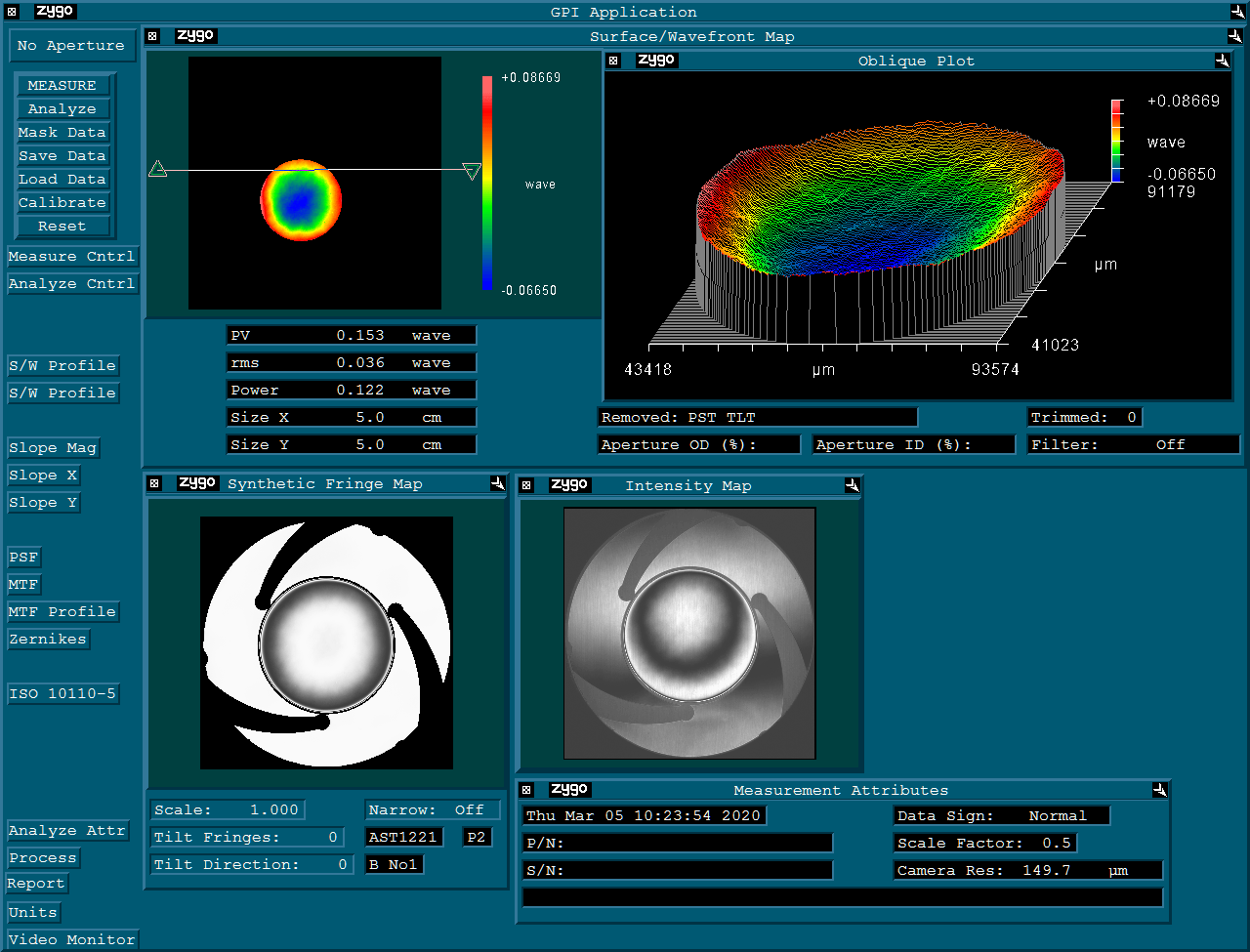}
\end{tabular}
\end{center}
\caption 
{ \label{fig:b1interf2}
The second interferometric and wavefront measurement of the B1 filter.} 
\end{figure}

\begin{figure}[!ht]
\begin{center}
\begin{tabular}{c}
\includegraphics[width=0.7\textwidth]{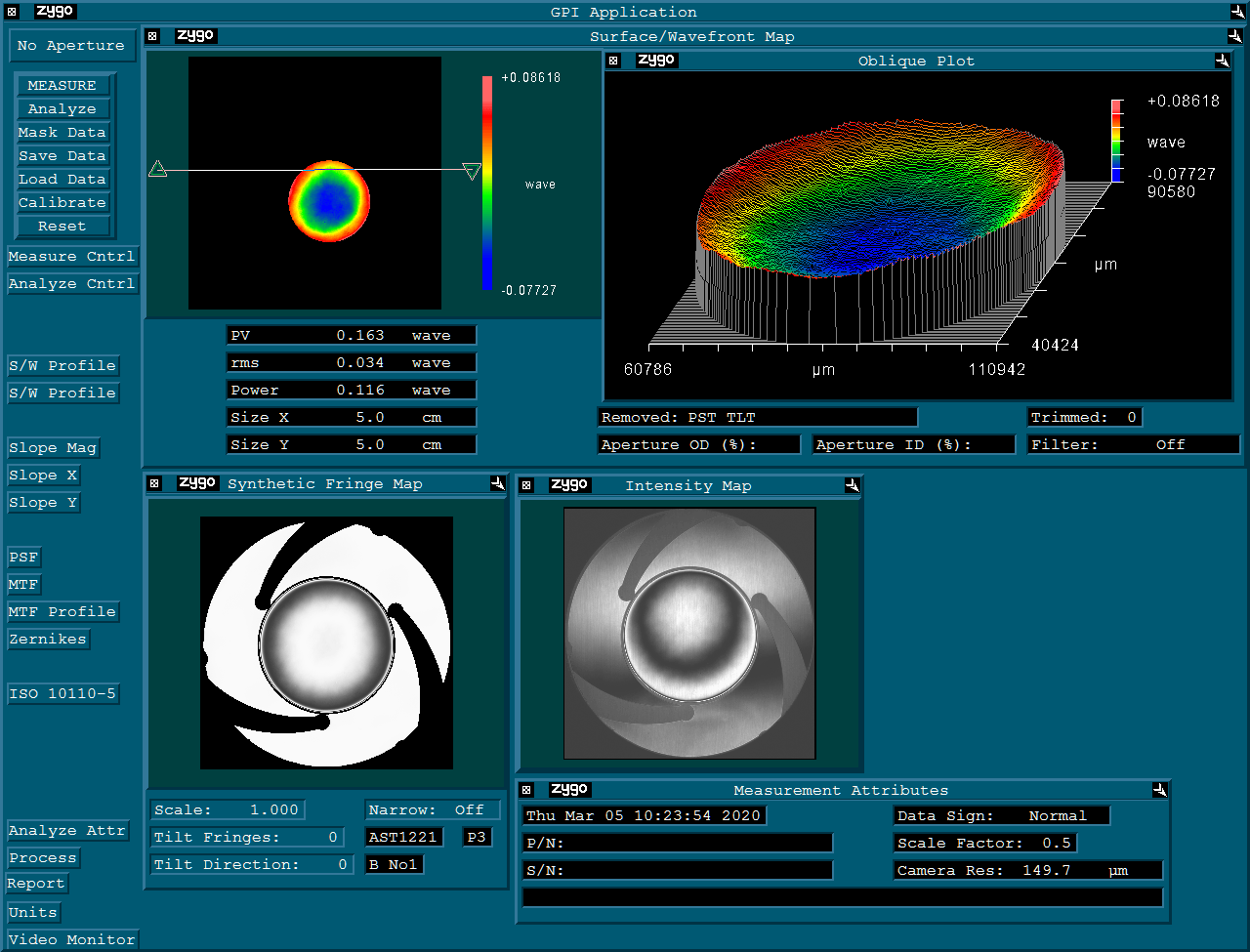}
\end{tabular}
\end{center}
\caption 
{ \label{fig:b1interf3}
The third interferometric and wavefront measurement of the B1 filter.} 
\end{figure}

The 3-level interferometry and wavefront measurement of the B1 filter is shown in Figures \ref{fig:b1interf1}, \ref{fig:b1interf2} and \ref{fig:b1interf3}.

\newpage

\subsection*{Disclosures}
The authors declare there are no financial interests, commercial affiliations, or other potential conflicts of interest that have influenced the objectivity of this research or the writing of this paper.

\subsection* {Code, Data, and Materials Availability} 
The data and archived version of the code presented in this article are publicly available at the GitHub repository: \url{https://github.com/HeisenbergK/NewWALOPCalibrate2Dens.git}.

\subsection* {Acknowledgments}
The PASIPHAE program is supported by grants from the European Research Council (ERC) under grant agreements No. 771282 and No. 772253; by the National Science Foundation (NSF) award AST-2109127;  by the National Research Foundation of South Africa under the National Equipment Programme; by the Stavros Niarchos Foundation under grant PASIPHAE; and by the Infosys Foundation.

VPa acknowledges support by the Hellenic Foundation for Research and Innovation (H.F.R.I.) under the “First Call for H.F.R.I. Research Projects to support Faculty members and Researchers and the procurement of high-cost research equipment grant” (Project 1552 CIRCE).

VPa acknowledge support from the Foundation for Research and Technology - Hellas Synergy Grants Program through project MagMASim, jointly implemented by the Institute of Astrophysics and the Institute of Applied and Computational Mathematics.

KT and AP acknowledge support from the Foundation for Research and Technology - Hellas Synergy Grants Program through project POLAR, jointly implemented by the Institute of Astrophysics and the Institute of Computer Science.

TG is grateful to the Inter-University Centre for Astronomy and Astrophysics (IUCAA), Pune, India for providing the Associateship programme under which part of this work was carried out.

VPe acknowledges funding from a Marie Curie Action of the European Union (grant agreement No. 101107047).


\bibliography{report}   
\bibliographystyle{spiejour}   


\vspace{2ex}\noindent\textbf{John Andrew Kypriotakis} is a Ph.D. student at the University of Crete, Greece, Department of Physics and at the Institute for Astrophysics of the Foundation for Research and Technology Hellas, Greece. He received his B.Sc. in Physics from University of Crete, Greece in 2017. He is currently working on the design and development of the WALOP instruments for the PASIPHAE survey. His areas of interest are Instrumentation (incl. Software), Data Analysis and Machine Learning.

\vspace{2ex}\noindent\textbf{Siddharth Maharana} is a post-doctoral researcher at the South African Astronomical Observatory. He received his Bachelor in Mechanical Engineering from Central University, Bilaspur, India in 2015 and his PhD in Astronomical Instrumentation in the Inter University Centre for Astronomy and Astrophysics, Pune India. He is currently working on the WALOP instruments for PASIPHAE survey. His areas of interest are polarimetric instrumentation and data analysis.

\vspace{1ex}
\noindent Biographies and photographs of the other authors are not available.

\listoffigures
\listoftables

\end{spacing}
\end{document}